\documentclass[12pt,letterpaper]{article}
\usepackage[a4paper, total={6.5in, 10in}]{geometry}
\usepackage{graphicx}
\usepackage{makecell}
\usepackage{authblk}
\usepackage{hyperref}
\usepackage{amsmath} 
\usepackage{amssymb} 
\usepackage{gensymb} 
\usepackage{afterpage}
\usepackage{orcidlink}
\usepackage{xcolor}
\usepackage[super,square,comma,sort&compress]  
   {natbib}
\usepackage{caption}
\newcommand*{\revision}[1]{\textcolor{black}{#1}}

\newcommand{\calB}{\mathcal{B}}%
\DeclareMathAlphabet\mathbfcal{OMS}{cmsy}{b}{n}
\newcommand{\bfB}{{\bf B}}%
\newcommand{\bff}{{\bf f}}\newcommand{\bfF}{{\bf F}}%
\newcommand{\bfn}{{\bf n}}%
\newcommand{\bfT}{{\bf T}}%
\newcommand{\bfu}{{\bf u}}%
\newcommand{\bfx}{{\bf x}}\newcommand{\bfX}{{\bf X}}%
\newcommand{\bfy}{{\bf y}}%
\newcommand{\bfz}{{\bf z}}%
\newcommand{\bfchi}{\boldsymbol{\chi}}%
\newfont{\tenbfit}{cmmib10}%
\newfont{\svnbfit}{cmmib8}%
\newfont{\tenbfsl}{cmbxti10}
\newfont{\mmit}{cmmi10}
\newfont{\smit}{cmmi9}
\newfont{\bfMit}{cmmi5}
\newfont{\tenbbb}{msbm10}%
\newfont{\svnbbb}{msbm8}%

\newcommand{\snb}{\boldsymbol{\mathsf{b}}}

\newcommand{\snh}{\boldsymbol{\mathsf{h}}}

\newcommand{\snm}{\boldsymbol{\mathsf{m}}}

\newfont{\tenssit}{cmssqi8 at 10pt}%
\newfont{\svnssit}{cmssqi8 at 7pt}%
\newfont{\gothic}{eufm10}%
\newfont{\sgothic}{eufm7}%
\newcommand{\trans}{{\mskip-2mu\scriptscriptstyle\top}} 
\newcommand{\inv}{{\scriptscriptstyle\mskip-1mu{-1}\mskip-2mu}}
\newcommand{\invtrans}{{\scriptscriptstyle\mskip-1mu{-\top}\mskip-2mu}}

\newcommand{\Blj}{\mbox{$\Big[\kern-0.275em\Big[$}}
\newcommand{\Brj}{\mbox{$\Big]\kern-0.275em\Big]$}}
\newcommand{\zed}{{\bf 0}}
\newcommand{\id}{{\bf 1}}

\newcommand{\Div}{\hbox{\rm Div}\mskip2mu}                                      
\newcommand{\Curl}{\hbox{\rm Curl}\mskip2mu}

\newcommand{\B}{{\calB}}
    
\newcommand{\X}{\bfX}
\newcommand{\x}{\bfx}

\newcommand{\F}{\bfF}

\newcommand{\FT}{\bfF^{\trans}}     	
	
\newcommand{\Def}{\overset{\text{def}}{=}}

\newcommand{\mat}{\text{\tiny R}}%
\newcommand{\T}{\bfT}

\newcommand{\Tmat}{\bfT_\mat}

\newcommand{\Gee}{\mbox{$G$}}

\newcommand{\Kay}{\mbox{$K$}}

\newcommand{\fluid}{\text{\tiny fluid}}

\newcommand{\Fit}{\bfF^{-\,\trans}}

\newcommand{\old}{{\text{old}}}

    \newcommand{\maxw}{\text{\tiny maxw}}

\newcommand{\Bbar}{\bar\bfB}
\newdimen\CdotAxis
\newcommand*{\CdotAux}[3]{%
  {%
    \settoheight\CdotAxis{$#2\vcenter{}$}%
    \sbox0{%
      \raisebox\CdotAxis{%
        \scalebox{#1}{%
          \raisebox{-\CdotAxis}{%
            $\mathsurround=0pt #2#3$%
          }%
        }%
      }%
    }%
    \dp0=0pt %
    \sbox2{$#2\bullet$}%
    \ifdim\ht2<\ht0 %
      \ht0=\ht2 %
    \fi
    \sbox2{$\mathsurround=0pt #2#3$}%
    \hbox to \wd2{\hss\usebox{0}\hss}%
  }%
}

\definecolor{purple}{rgb}{.63,.36,.94}
\definecolor{verdigris}{rgb}{0.26, 0.7, 0.68}
\definecolor{MITgray}{cmyk}{0 0 0 0.1}

\definecolor{sgc1}{rgb}{.5 .5 0}
\definecolor{sgc2}{rgb}{1 .4 0}

\definecolor{sgc1}{rgb}{1 0 0}
\definecolor{sgc2}{rgb}{0 0 1}

\makeatletter
\renewcommand{\maketitle}{\bgroup\setlength{\parindent}{0pt}
\begin{flushleft}
  \textbf{\@title}
  
  \@author
\end{flushleft}\egroup}
\makeatother

\title{Magnetically Responsive Microprintable Soft Nanocomposites with Tunable Nanoparticle Loading}

\date{}

\author[1,\orcidlink{0000-0001-6396-1720}]{Rachel M. Sun}
\author[1,\orcidlink{0000-0002-4340-7017}]{Andrew Y. Chen}
\author[2,\orcidlink{0000-0003-0891-6462}]{Yiming Ji}
\author[3,\orcidlink{0000-0002-4572-004X}]{Eric M. Stewart}
\author[2,\orcidlink{0000-0002-4114-6167}]{Daryl W. Yee}
\author[1,*,$\dagger$,\orcidlink{0000-0002-2649-4235}]{Carlos M. Portela}

\affil[1]{Department of Mechanical Engineering, Massachusetts Institute of Technology, Cambridge, MA 02139, USA}
\affil[2]{Institute of Electrical and Micro Engineering, EPFL, Lausanne, CH-1015, Switzerland}
\affil[3]{Department of Mechanical and Materials Engineering, University of Cincinnati, Cincinnati, OH 45221, USA}
\affil[*]{Correspondence: cportela@mit.edu}
\affil[$\dagger$]{Lead Contact: cportela@mit.edu}

\begin{document}

\maketitle

\section*{Summary}
Magnetic remote actuation of soft materials is attractive for applications such as transforming materials and medical robots. However, due to manufacturing limitations, microscale magnetoactive devices are scarce---light-based additive manufacturing methods, despite achieving microscale resolution, struggle with particle-induced light scattering. Moreover, large hard-magnetic microparticles restrict ultimate feature sizes, and deformation of soft-magnetic nanoparticle composites requires impractically high loading and field gradients. Among successfully fabricated microscale soft-magnetic composites, limited control over particle loading, distribution, and matrix-phase stiffness has hindered their functionality. Here, we combine two-photon polymerization with iron oxide nanoparticle coprecipitation to fabricate 3D-printed microscale nanocomposites with spatially tunable nanoparticle distribution. We control nanoparticle content by locally modulating the two-photon dose, imbuing parts with varied magnetic functionality and achieving millimeter-scale elastic deformations, demonstrated by a soft robotic gripper and a bistable bit register and sensor. Our approach enables precise control of mechanical and magnetic properties towards microscale metamaterial and robotics applications.

\subsection*{Keywords}
magnetic metamaterials, stimuli-responsive, nanocomposites, two-photon polymerization 

\newpage

\section{Introduction} 

Stimuli-responsive materials with time-dependent responses have recently been a subject of significant research interest. These materials respond to different types of stimuli including thermomechanical\cite{ji20214d}, mechanical\cite{haghpanah2016multistable}, piezoelectric\cite{cui2022design}, and electromagnetic\cite{jackson2018field, Chen2021}, among others\cite{dong2022untethered, smart2024magnetically}. However, invoking a stimulus response in these materials often involves sensitive chemical processes, with extended time scales, or require direct physical contact. In contrast, magnetically actuated materials respond nearly instantaneously and can be actuated at a distance. As a result, magnetically responsive materials have attracted interest for applications in soft robotics\cite{ze2020magnetic,lee20203d,carpenter2021facile}, medical devices \cite{kim2019ferromagnetic, cabanach2020zwitterionic}, and actuatable microsystems \cite{ni2021core, liu2022magnetically, cui2019nanomagnetic,Chen2024}. 

Magnetic remote actuation of two-dimensional (2D) and three-dimensional (3D) structures has been realized at the macroscale (milli- to centimeter scales) in a variety of ways. One popular approach is to use magnetic particles embedded in a compliant matrix material to create magnetorheological elastomers (MREs). Commonly, hard-magnetic particles---i.e., particles with high remanent magnetization such as neodymium iron boron (NdFeB)---are used to create so-called hard-magnetic MREs. These MREs demonstrate a wide range of applications across soft transforming materials\cite{kim2018printing, wu2022magnetically,zhang2023magnetic,sim2025selective,song2020reprogrammable}, reconfigurable devices and robots\cite{zhang2021voxelated,sun2022magnetic,ze2022spinning,ma2020magnetic}, and soft continuum medical devices\cite{kim2019ferromagnetic, wang2022adaptive}. These methods typically use direct ink writing \cite{kim2018printing, CrespoMiguel2025}, molding techniques \cite{alapan_reprogrammable_nodate,PerezGarcia2025,MorenoMateos2022}, or magnetically-assisted vat photopolymerization \cite{Wu2023} to align hard-magnetic particles as the matrix material solidifies; in a uniform magnetic field, this alignment is essential to the actuation of hard-magnetic MREs under torque-induced deformation\cite{kim2022magnetic}. Moreover, in addition to realizing visible elastic deformations, the material properties (e.g., stiffness) of these hard-magnetic MREs can change with the applied field. Since macroscale fabrication techniques can readily tune particle loading and orientation, it is possible to architect structures with highly tunable properties after fabrication. 

However, at the microscale, there are drastically fewer realizations of magnetically active devices\cite{liu2022magnetically} and microrobots\cite{cabanach2020zwitterionic, giltinan20213d, ceylan20183d,kim2016fabrication}. Because the minimum size of commonly available hard-magnetic particles (approximately tens of micrometers) exceeds typical microscale feature sizes ($\sim$10 \textmu{}m), multi-step processing is typically required to overcome the corresponding fabrication limitations when they are used \cite{Li2024, Rothermel2024, smart2024magnetically, giltinan20213d}. Typically, single-domain soft-magnetic particles of $\sim$200 nm in size, with low remanent magnetization, are used instead. In contrast to macroscale hard-magnetic MREs, microrobots and other microscale devices with soft-magnetic particles are actuated in spatially varying magnetic field gradients by gradient pulling, often by a rotating magnetic field or permanent magnet. Although isotropic soft-magnetic MRE behavior is well-characterized theoretically\cite{Mukherjee2020, stewart2025magnetostriction,han2013field,Danas2012}, there are few examples of soft-magnetic materials with microscale features that demonstrate elastic body deformations\cite{cestarollo2022nanoparticle}; the majority of soft-magnetic MREs undergo magnetostriction or other rigid-body motions. This lack of visible elastic deformation is largely due to weak gradient fields, low particle loading, or a stiff matrix phase in low-aspect-ratio geometries.

One family of fabrication routes that has successfully yielded 3D deformable materials at the microscale is molding\cite{Bodelot2017,cestarollo2022nanoparticle, liu2022magnetically, alapan_reprogrammable_nodate,Moreno2021,GonzalezSaiz2025}. However, these methods rely on mold-release steps or the existence of pre-coated layers of material on a substrate, which limits the landscape of realizable geometries, particularly in the case of non-convex structures. Moreover, while these methods allow for particle alignment during fabrication, they do not enable spatial modulation of nanoparticle distribution. On the other hand, the recent advent of commercially-available high-throughput two-photon polymerization (TPP) 3D-printing systems enables freeform fabrication of arbitrary geometries. In particular, TPP can be leveraged to print structures from a resin containing functionalized particles, commonly iron oxide nanoparticles (IONPs)\cite{cabanach2020zwitterionic, ceylan20183d}, or to perform low-power two-photon crosslinking on pre-polymerized or pre-baked layers containing IONPs\cite{geid2024gym, peters2014superparamagnetic}. However, applications are hindered by limited control over the subsequent nanoparticle distribution, and printable structure sizes are limited due to sedimentation and scattering effects during printing. In many cases, the IONP concentration required to produce measurable deformation also induces light scattering, which precludes printing altogether. Other methods have tried to overcome these limitations through multimaterial printing and subsequent attachment of magnetic components to non-magnetic components\cite{soreni2020multimaterial} and \emph{in situ} particle growth after printing\cite{xing2025trimag}. To date, fabrication methods for microscale magnetically responsive materials and devices lack the ability to simultaneously program microscale geometry and spatial nanoparticle loading---two key parameters for achieving programmable, large visible elastic deformations at the microscale. 

In this work, we demonstrate a robust, three-dimensional freeform manufacturing method for magnetically responsive microscale parts with a tunable loading of magnetoactive material. Our method is based on additive manufacturing followed by an infusion-precipitation sequence which has been successfully demonstrated using macroscopic vat-photopolymerization techniques\cite{Ji2025}. We extend this concept to microscale 3D-printing by utilizing two-photon polymerization, which further enables precise three-dimensional spatial modulation of the incident dose. We leverage this tunability and demonstrate that as we vary the dose, the concentration and distribution of resulting iron oxide nanoparticles (IONPs) in the composite changes in a controllable manner. With this spatial modulation, we characterize the magnetic and mechanical properties of the IONP nanocomposite, and demonstrate large (i.e., optically measurable) magnetically-induced component deformations which can be programmed to vary within the same printed structure. We demonstrate the capabilities of our fabrication method by printing and magnetically actuating a microscale gripper and a bistable structure bit with applications in encryption and sensing. \revision{Using a coupled magneto-elastic model and the finite element method, we numerically analyze the behavior of the IONP nanocomposite, quantitatively linking process parameters during fabrication to the response of these composite structures to an applied external field.} Altogether, this approach enables precise control of IONP content and mechanical properties to fabricate parts with tunable magnetic responsiveness and structural integrity at the microscale, advancing magnetically responsive microscale devices and materials toward applications in medical devices and microscale soft robotics. 

\section{Results}
\subsection{Microscale magnetic nanocomposite fabrication}
We fabricate magnetically active iron oxide nanoparticle (IONP) nanocomposites using a two-step process. In the first step, we fabricate a poly(ethylene glycol) diacrylate (PEGDA)-based hydrogel precursor using two-photon polymerization (TPP) 3D-printing (UpNano NanoOne). The use of TPP enables the freeform fabrication of microscale three-dimensional geometries, while maintaining spatial control of the incident two-photon dose and hence crosslink density in the hydrogel. After printing, the PEGDA hydrogel is immersed in an aqueous FeCl$_2$/FeCl$_3$ solution to introduce iron ions, followed by ammonium hydroxide to synthesize IONPs \emph{in situ} via ammonia-induced coprecipitation of Fe\textsuperscript{2+} and Fe\textsuperscript{3+} salts (Figure 1A). \revision{Specifically, the ammonia reacts with water to generate hydroxide ions that react with the infused Fe$^{2+}$ and Fe$^{3+}$ ions to form iron hydroxides, which then aggregate to form iron oxide\cite{Hong2007, xing2025trimag}.}
\begin{figure}[hbt!]
\begin{centering}
\includegraphics[width=0.9\textwidth]{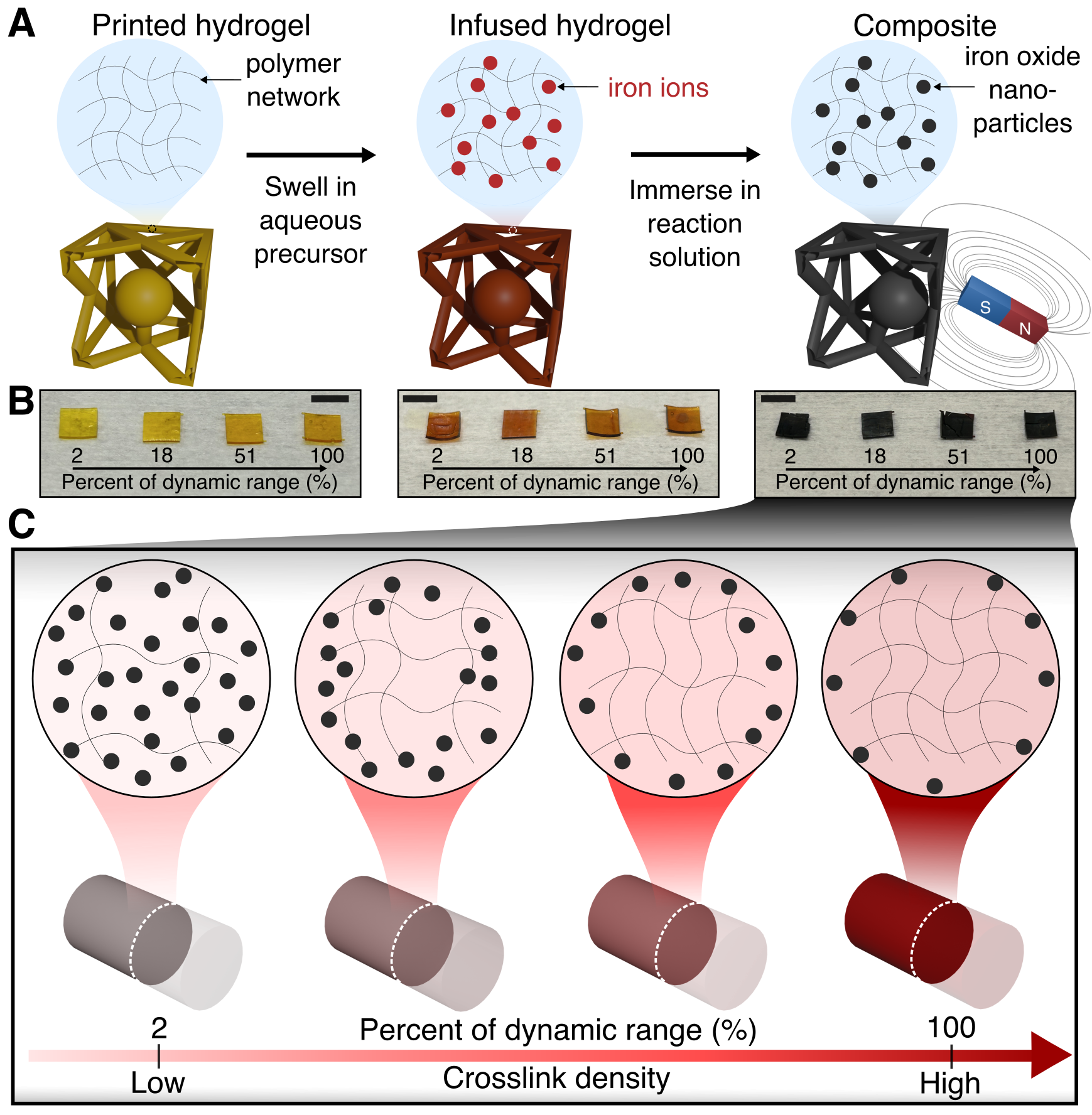}
\caption{\textbf{Fabricating magnetically responsive nanoparticle composites.} (A) The fabrication process involves printing a hydrogel via two-photon polymerization (TPP), infusing with iron ions in an iron salt bath, then immersing the sample in ammonium hydroxide to coprecipitate IONPs within them. (B) Images of printed samples as a function of laser power (or dose) after each stage of the fabrication process. Scale bars, 5 mm. (C) Increasing the two-photon dose during printing creates higher polymer crosslink density, resulting in lower amounts of iron ion diffusion and subsequent nanoparticle content, corresponding to a change in the quantity and spatial distribution of magnetic nanoparticles.
}
\label{fig:1overview}
\end{centering}
\end{figure}
By spatially modulating the incident laser power in the TPP step, we can produce parts with varying degrees of PEGDA crosslink density. Indeed, we noticed darkening color of the printed samples with increasing dose (Figure 1B). In two-photon polymerization, the dose $D$ is related to the incident laser power $I$ and the laser scanning speed $v$ by 
\begin{equation}
D \propto \frac{I^2}{v},
\label{eq:dose}
\end{equation}
allowing control of physical and mechanical properties of the crosslinked polymer by varying the incident dose via laser power and scanning speed modification\cite{Bauer2019, Eren2025}. In our case, the dose affects the extent to which iron ions and ammonia solution can diffuse into the bulk of the hydrogel during the subsequent immersion steps (Figure 1C). In turn, by varying the dose we can spatially program the concentration and spatial distribution of IONPs in the resulting composite, which directly affects its response to an external field gradient. To make these trends broadly applicable to various TPP systems, we express the incident dose as a percentage of the dynamic range, which represents the range of two-photon dose consistent with polymerization. Under this system, 0\% corresponds to the minimum polymerization threshold dose, whereas 100\% corresponds to the onset of overpolymerization-induced cavitation. \revision{Since at fixed scanning speed---chosen to maximize throughput---dose increases quadratically with increasing incident laser power $I$, we express dynamic range $R$ such that 
\begin{equation}
    R = a(I-I_{\text{min}})^2
    \label{eq:dynamicRange}
\end{equation}
where $I_{\text{min}}$ is the minimum incident laser power to yield polymerization and $a$ is a fitting constant calculated by fitting Equation \ref{eq:dynamicRange} to end points $(I = I_{\text{min}}, R = 0)$ and $(I = I_{\text{max}}, R = 1)$ where $I_{\text{max}}$ is the onset of overpolymerization-induced cavitation (see SI section A2 and Table S4 for best-fit parameters). 
}

\subsection{Tuning nanoparticle distribution and magnetic properties}
To understand the effect of crosslink density on the extent of diffusion during this \emph{in situ} ammonia-induced coprecipitation process, we fabricated four monolithic IONP composite cylinders of diameter 1 mm and height 1 mm, each printed at a different dose. We controlled the dose across samples by changing the laser power, maintaining a globally constant laser scanning speed and repetition rate for each (Section~\ref{sec:printing}). Corresponding to this range of laser powers, our resin formulation allows a six-fold change in the dose between the lower and upper limits of the dynamic range. After printing and coprecipitation, we sliced each cylinder across its diameter, such that energy-dispersive X-ray spectroscopy (EDS) measurements (Oxford Ultim Max) could be performed on at least half of a flat cross section \revision{(Figure S1)}. We obtained line scans of the cross sections, normalizing each scan by its maximum number of counts to allow comparison across all scans (Section~\ref{sec:eds}).  

Line scans across slices of varying polymerization dose elucidate the effect of crosslink density on resulting IONP distribution (Figure 2). All cross-sections demonstrate various extents of polymer-rich core and iron-rich shell sections, which we refer to as the ``core-shell effect." The 1\% dynamic-range cross-section is the most spatially uniform, having a negligible core section compared to the cross-sections of higher crosslink density. The total iron content and shell thickness decrease drastically with increasing crosslink density, as the line scans of the 13\% to 98\% dynamic-range cross-sections show significantly decreased iron counts in the shell region compared to the 1\% dynamic range cross-section. Characterization via SEM confirms this core-shell effect, demonstrating \revision{distinct} core and shell regions, as shown in the 13\% dynamic-range sample (Figure 2A, panel ii). 

\begin{figure}[hbt!]
\begin{centering}
\includegraphics[width=0.9\textwidth]{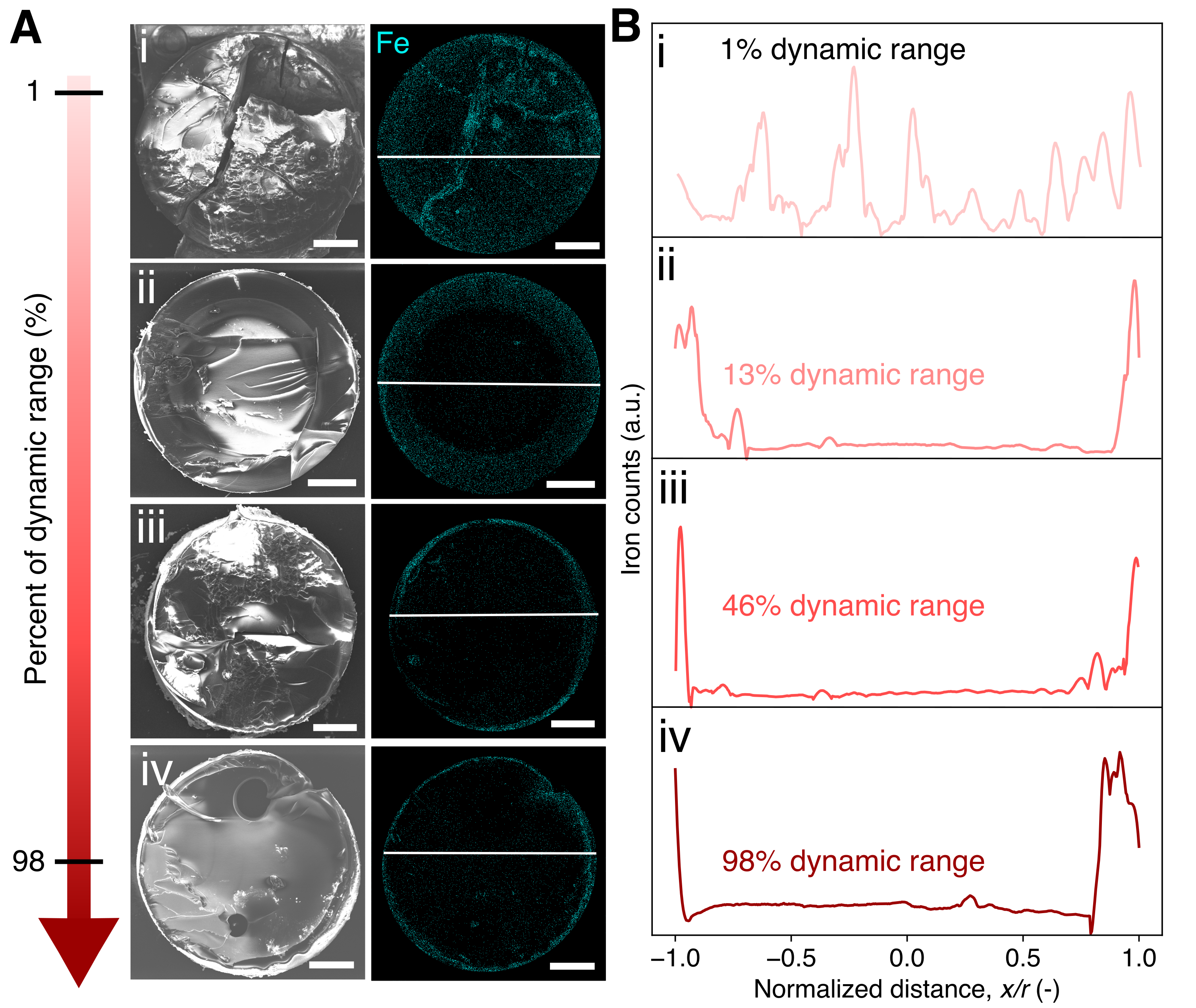}
\caption{\textbf{Energy dispersive x-ray spectroscopy (EDS) characterization of the magnetic composite.} (A) Scanning electron microscope images (left) and corresponding EDS area scans of iron (Fe) counts (right) as a function of the dynamic range. (B) EDS line scans of cross-sections of an IONP composite block printed at \revision{(i) 1\%, (ii) 13\%, (iii) 46\%, and (iv) 98\% dynamic range}. Normalized distance $x/r$ represents the line scan location $x$ along the radial distance $r$ of the white lines in (A). Scale bars, 200~\textmu{}m. Details on EDS map data processing are provided in Figure S1.} 
\label{fig:2eds}
\end{centering}
\end{figure}

\revision{To understand the physical origin of the core-shell effect, we conducted an intermediate set of EDS measurements, taken after ion infusion but before coprecipitation (Figure S2). Across the full range of crosslink densities, we observed full, homogeneous infiltration of iron ions throughout the 1 mm diameter cross-section. This suggests that the immersion step is sufficient to establish a spatially uniform precursor distribution of iron ions regardless of crosslink density.}

\revision{Moreover, with the ion infusion time fixed, we measured areal EDS maps of the cylinder cross-sections for varying coprecipitation (i.e., ammonium hydroxide immersion) times (Figure S3). As the coprecipitation time increases, we observe the development of the IONP-rich shell phase from the outside in, consistent with the concentration gradient of hydroxide ions originating from the ammonia. To quantify the extent of the core-shell development as a function of crosslink density in the polymer, we measured a characteristic diffusion coefficient, $D_\mathrm{eff}$, for which the shell thickness, $L$ (taken to be a characteristic length scale of diffusion), is related to the diffusion time, $t$, by 
\begin{equation}
    L = 2(D_\mathrm{eff} t)^{1/2}.
\end{equation}
The measured characteristic diffusion coefficients (Figure S4) show over an order-of-magnitude difference between the lowest crosslink density (1\% dynamic range, $D_\mathrm{eff} = 55.2$ \textmu{}m$^{2}$ s$^{-1}$) and an intermediate value (46\% dynamic range, $D_\mathrm{eff} = 2.2$ \textmu{}m$^{2}$ s$^{-1}$), suggesting that the density of the polymer network heavily modulates the extent to which IONPs can form within the composite. In particular, a higher crosslink density suppresses diffusion into the center of the bulk hydrogel during ammonium hydroxide infusion, restricting the formation of IONP to a thin shell near the perimeter of the part.} Broadly, given these measured effective diffusion coefficients, the process time and geometry can be balanced to control the extent to which a printed cross-section experiences complete infiltration. To demonstrate this, we measured the IONP distribution in cross-sections of cylinders of different diameters but otherwise processed identically, confirming that below a critical diameter, the entire cross-section experiences complete infiltration (Figure S6).

\revision{Building on the established process-structure relation and the ability of the TPP system to spatially modulate the two-photon dose, we fabricated a heterogeneous cylinder consisting of four printed-in-place quarter sections, each programmed with a distinct crosslink density (Figure S5). Areal EDS mapping of a cross-section, together with appropriate radial line scans, confirms that each region exhibits a distinct degree of core-shell formation, in agreement with the trends observed in homogeneous cylinders, despite being part of a single contiguous structure.}

Characterization of the resulting magnetic properties of the nanocomposite further confirms this trend of increasing IONP with decreasing crosslink density. We fabricated and dried IONP nanocomposite samples of different laser powers with nominal dimensions of 5$\times$5$\times$0.5 mm\textsuperscript{3} for characterization using vibrating sample magnetometry (VSM) through the 0.5 mm sample thickness to obtain their magnetization curves (Figure 3A). These curves demonstrate that the 2\% dynamic range sample exhibits a higher saturation magnetization than the other high-crosslink-density samples, and thus higher responsiveness to magnetic forces (Table S2). \revision{We also performed VSM measurements on identical, hydrated samples (Figure S7A), which display the same behavior as their dry counterparts owing to the negligible magnetic susceptibility of water.} 

X-ray diffraction (XRD, Malvern Panalytical Empyrean) confirms the presence of iron oxide in the composites (Figure S8), as well as the magnetization trends observed in the VSM measurements. Characteristic peaks corresponding to cubic spinel iron oxide (Fe$_{3}$O$_4$)\cite{Laurent2008} are clearly present in the lowest crosslink density specimen at $13.7\degree$ (220), $16.1\degree$ (331), $19.5\degree$ (400), $25.5\degree$ (511), and $27.7\degree$ (440). As the crosslink density increases, the sharpness and prominence of the peaks decreases, indicating decreasing crystallinity of the iron oxide spinel structure, consistent with our other measurements of the concentration of IONPs (Figure 3A). 

\subsection{Mechanical properties of the nanocomposite}
As a result of the varying crosslink density and the core-shell effect, the change in mass of the samples varies as a function of incident laser power. We quantified this change in mass by weighing block-shaped specimens of nominal dimensions 1.4$\times$1.4$\times$1 mm$^3$ before and after coprecipitation (Figure S9). The 2\% dynamic-range samples, which had the lowest crosslink density, had the greatest average mass change of 32\% compared to 10\% for the highest crosslink density samples. This corroborates our findings above which suggest that these samples allowed the greatest amount of IONP diffusion and growth. Moreover, as laser power increases (and the core-shell effect grows stronger), the mass change decreases (Table S2). This suggests that the trends in magnetization seen previously are a result of a combination of the distribution (core-shell effect) as well as the concentration of IONPs.

Mechanical properties also reflect the change in crosslink density with varying incident dose. We conducted uniaxial compression experiments (Figure 3B) and stress-relaxation experiments (Figure S10) with a nanoindenter (Alemnis ASA) on dried rectangular pillars with nominal dimensions of 100$\times$100$\times$300 \textmu{}m\textsuperscript{3}. As the incident dose increased, the stiffness of the composite increased, tending toward a saturated stiffness of 53.2 MPa as the dose approached 100\% dynamic range. \revision{Through a strain of 15\%, pillars across the dynamic range exhibited minimal deviation from linearity and recovered nearly completely upon unloading (Figure 3B, inset).} These trends are consistent with overall patterns observed in polymers fabricated using TPP in which the mechanical properties and degree of conversion both increase with dose until saturation \cite{Jiang2014, Bauer2019}. \revision{In the hydrated state, the composite exhibits higher compliance (with a measured stiffness of 32.2 MPa at 98\% dynamic range) but similar overall behavior (Figure S7B).} The measured Poisson's ratio (Figure S11) remains approximately constant with laser power, indicating it is not affected by the presence of the core-shell structure.

\begin{figure}[h]
\begin{centering}
\includegraphics[width=\textwidth]{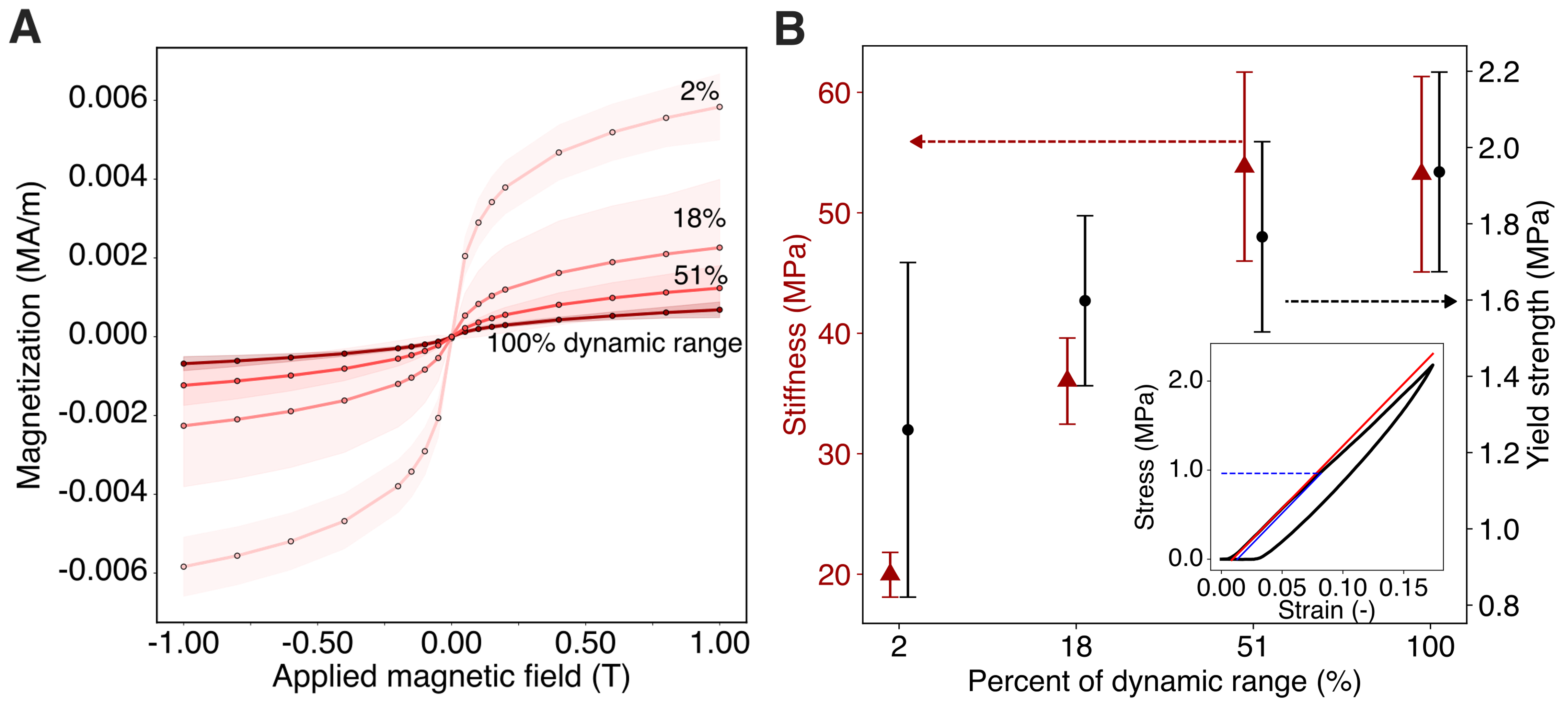}
\caption{\textbf{Magnetic and mechanical properties of the \revision{dried} IONP composite material.} \small{
(A) Magnetization curves of monolithic 5 $\times$ 5 $\times$ 0.5 mm\textsuperscript{3} blocks printed with different laser powers: 2\% (light pink), 18\%, 51\%, and 100\% (red) dynamic range. Shaded regions indicate one standard deviation from the mean (solid lines) across three sample replicates. 
(B) \revision{Stiffness (red triangles) is plotted against percent of dynamic range.} An example stress-strain curve with stiffness obtained from the linear-loading-regime slope (red line) are shown in the inset. }}
\label{fig:3properties}
\end{centering}
\end{figure}

\afterpage{
\begin{figure}[p]
\begin{centering}
\includegraphics[width=0.7\textwidth]{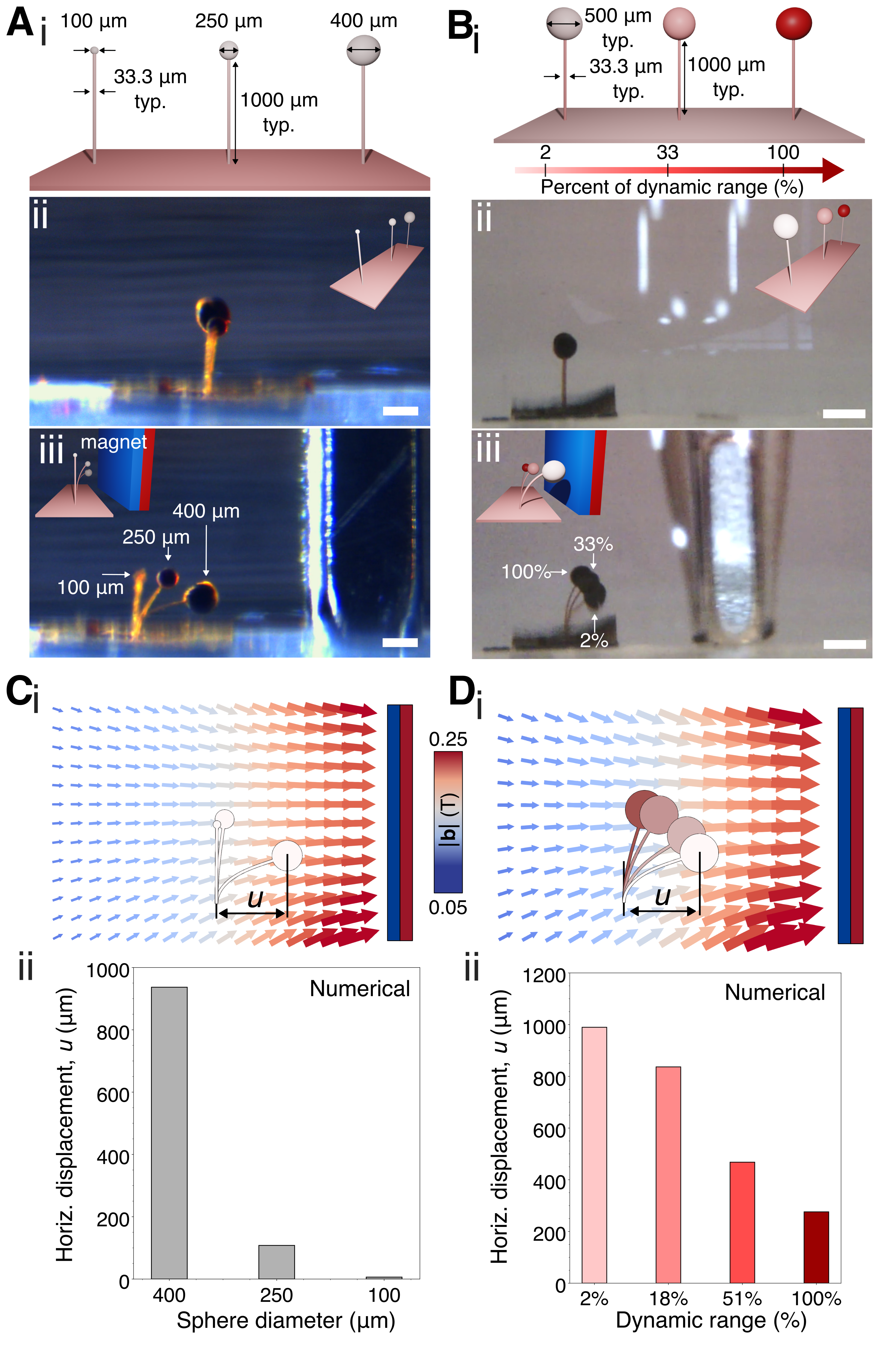}
\caption{\revision{\textbf{Deflection response of simple sphere arrays.}
(A) Schematic (i) and experimental demonstration (ii)-(iii) of the deflection of an array of spheres of different diameters printed with constant 2\% dynamic range, attached to cylinders printed at 18\% dynamic range and a base printed at 51\% dynamic range. Images show the array from a side view before and after a permanent magnet is placed near enough to deflect the array. Scale bars, 400~\textmu{}m. 
(B) Schematic (i) and experimental demonstration (ii)-(iii) of the deflection of an array of spheres of varying laser powers, attached to cylinders printed at 18\% dynamic range and a base printed at 51\% dynamic range. Images show the array from a side view before and after a permanent magnet is placed near enough to deflect the array. Scale bars, 1 mm. 
(C) (i) Numerical prediction of horizontal displacements $u$ of spheres of diameters 100~\textmu{}m, 250~\textmu{}m, and 400~\textmu{}m in response to a 30 mT mm$^{-1}$ field gradient induced by a magnet 2 mm away. (ii) Numerical measurements of horizontal displacement $u$ of different diameter spheres away from the initial vertical axis are plotted for different sphere diameters.  
(D) (i) Numerical prediction of horizontal displacements $u$ of spheres with dynamic ranges 2\% (light pink), 18\% (pink), 51\% (dark pink), and 100\% (red) in response to a 30 mT mm$^{-1}$ field gradient induced by a magnet 2 mm away. (ii) Numerical measurements of horizontal displacement $u$ of the spheres are plotted for different dynamic range percents. } 
\\\hspace{\textwidth} 
}
\label{fig:4ADeflection}
\end{centering}
\end{figure}
}

\subsection{Towards magnetically active micro-components}
To demonstrate the potential of our processing route to develop functional microscale components, we fabricated prototypes of directly printable magnetically responsive structures which utilize the ability to spatially modulate the incident laser dose and thus tune the magnetic response. 

\subsubsection{Deflection response of simple sphere arrays} To evaluate the combined influence of laser-power-dependent magnetic and mechanical properties, we fabricated composite samples featuring arrays of spheres on cylindrical pillars (Figuer 4A, i), which illustrate how variations in crosslink density affect the net response of a magnetically responsive structure. To understand the effect of iron oxide fill fraction 
needed to deflect a cylinder of 33.3~\textmu{}m, we fabricated spheres of sizes 100~\textmu{}m, 250~\textmu{}m, and 400~\textmu{}m and placed them in the presence of a 2.54 $\times$ 0.64 $\times$ 0.32 cm\textsuperscript{3} BX042-N52 permanent magnet (K\&J Magnetics, Inc.). The 100~\textmu{}m sphere did not cause any optically observable deflection of the cylinder while the 250~\textmu{}m and 400~\textmu{}m spheres caused increasing levels of deflection of their respective pillars (Figure 4A, panels ii-iii; Supplemental Video 1). A similar effect occurred for spheres printed with different laser powers with the sphere of 2\% dynamic range causing the largest deflection (Figure 4B; Supplemental Video 2).  

\revision{\subsubsection{Coupled magneto-elastic modeling of magnetically actuated components}} \revision{To quantitatively analyze the behavior of our IONP composite under an applied external field gradient, we used a finite-element implementation of the coupled theory from Ref.\cite{stewart2025magnetostriction} (summarized in SI section A9). We model the magnetizable solid body, endowed with suitable magnetic and mechanical properties derived from experiments (Table S2), and the surrounding fluid (air) domain, which is necessary to capture the changing magnetic fields surrounding the solid. For simplicity we use a two-dimensional formulation, adjusting the total magnetization in each magnetizable body appropriately to match the three-dimensional geometries used in experiments. Upon the application of suitable mechanical boundary conditions and the magnetic field gradients associated with the bar magnets used in experiments (Table S5), magnetic Maxwell tractions acting at the surface of the magnetizable body drive large elasto-dynamic deformations of the body through the fluid surroundings. Dynamic effects are modeled using an implicit Newmark time-integration scheme and a viscous body force term that accounts for hydrodynamic drag and provides numerical damping. A single sphere on a cylinder is modeled, demonstrating a change in magnetization direction and magnitude within the sphere in response to an approaching magnet as the sphere deflects the cylinder towards the magnet (Supplemental Video 3). Additional details regarding the finite element modeling approach are given in SI section A10.}

\revision{Using this approach, we simulated the response of the simple sphere arrays of varying size (Figure 4C, panel i) and of varying laser power (Figure 4D, panel i). In each case, the equilibrium deflections show good qualitative agreement with the experimental photographs, confirming that in the presence of a field gradient, increasing the total magnetization of the sphere either by increasing the structure size or by lowering the crosslink density results in a larger deformation magnitude. We quantified the horizontal displacement, $u$, of each sphere, demonstrating a 150-fold difference in displacement between the smallest and largest spheres at constant crosslink density (Figure 4C, panel ii), and a 3.5-fold difference in displacement between the least and most crosslinked sphere at constant size (Figure 4D, panel ii).} \\

\subsubsection{Microscale magnetically actuated gripper}
To demonstrate the facility with which our IONP-based structures can be remotely actuated, as well as the potential to spatially tune the magnetic response by changing the crosslink density, we fabricated a microscale gripper (\revision{Figure 5A-B}; Supplemental Video 4). The gripper consisted of equal-sized spheres connected by thin arms to a center post. Two of the spheres were printed at 2\% dynamic range, corresponding to the magnetically strongest parameters. The other two spheres were printed at 51\% dynamic range, resulting in a less magnetically responsive structure. The center post and arms were printed at a constant 18\% dynamic range. When a magnet was introduced into the proximity of the gripper, the arms bent upwards asymmetrically, consistent with the difference in their dose (\revision{Figure 5B}, panels ii-iii). 

\begin{figure}[h]
\begin{centering}
\includegraphics[width=0.9\textwidth]{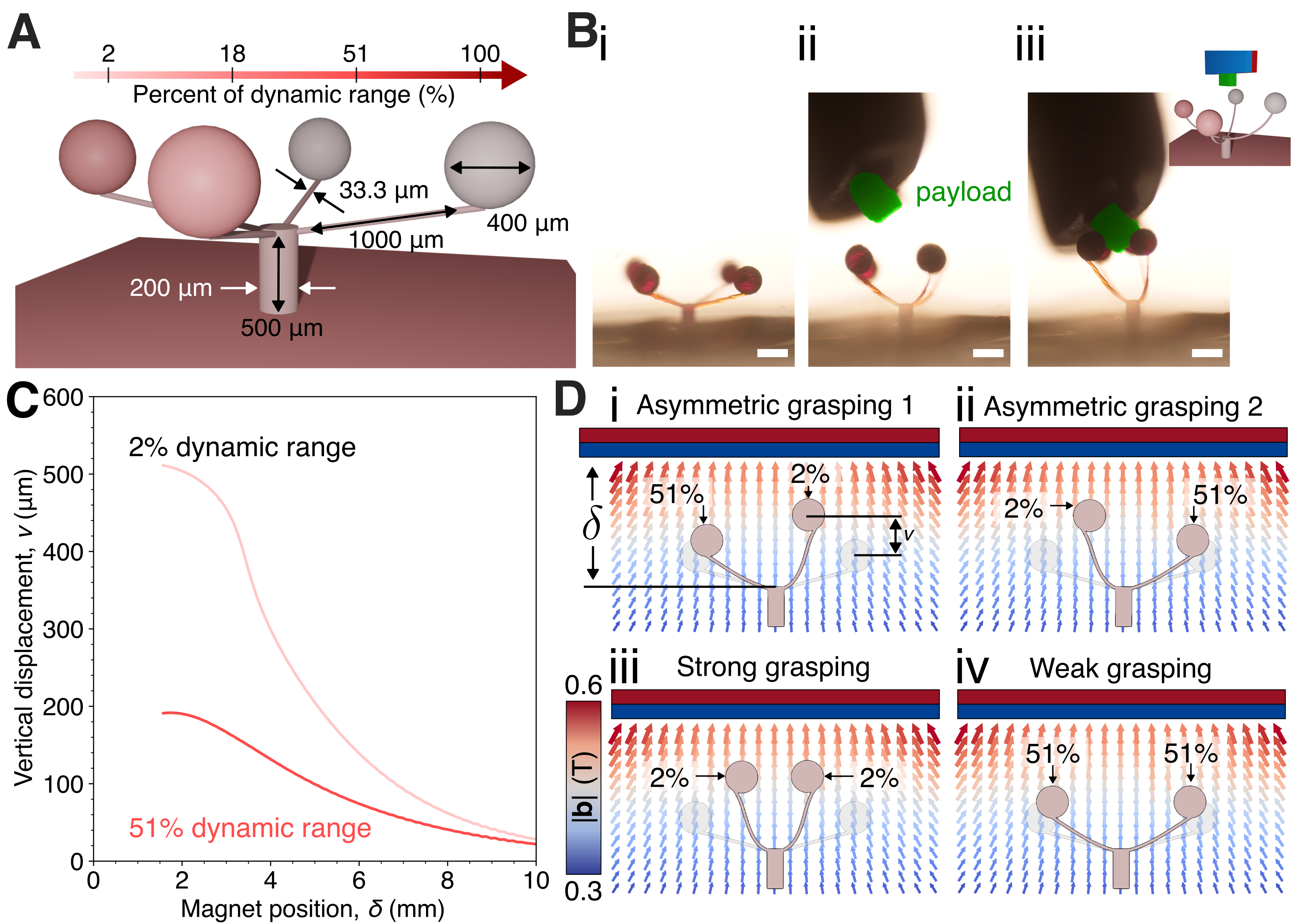}
\caption{\revision{\textbf{Functional microscale gripper via IONP nanocomposite.}}
(A) A microscale gripper is fabricated with varying two-photon doses, resulting in a magnetically-active state with varying degrees of deflection of different arms.
(B) Experimental microscope images show a gripper holding onto a payload attached to the permanent magnet (green). Scale bars, 400~\textmu{}m. 
\revision{(C) Vertical displacement $v$ of the 2\% (light pink) and 51\% (dark pink) dynamic range arms are displayed as a function of magnet position $\delta$.
(D) 2D numerical models of (i) right-arm-dominant asymmetric grasping, (ii) left-arm-dominant asymmetric grasping, (iii) strong grasping, and (iv) weak grasping in response to a 120 mT mm$^{-1}$ field gradient induced by a magnet 2 mm away. 
} 
\\\hspace{\textwidth} 
}
\label{fig:4Bgripper}
\end{centering}
\end{figure}

\revision{Numerically predicting the vertical displacement $v$ of each gripper arm as a function of magnet position $\delta$ reveals strongly contrasting behavior between a highly magnetized arm (printed at 2\% dynamic range), and a more weakly magnetized arm, printed at 51\% dynamic range (Figure 5C). We can further extend this result to develop various functional grasping modes (Figure 5D), including two configurations of asymmetric grasping, strong grasping, and weak grasping. These demonstrations represent one way in which varying crosslink density can be used to encode distinct actuation modalities into each arm. Encoding arms with dynamic ranges between 2\% and 51\% can enable a wider variety of vertical displacements $v$ in response to a magnet, representing a pathway toward highly tunable functional behaviors. Such spatial programmability provides a route towards multifunctional soft robotic systems capable of adaptive, field-controlled remote manipulation.}

\afterpage{
\begin{figure}[p]
\begin{centering}
\includegraphics[width=0.9\textwidth]{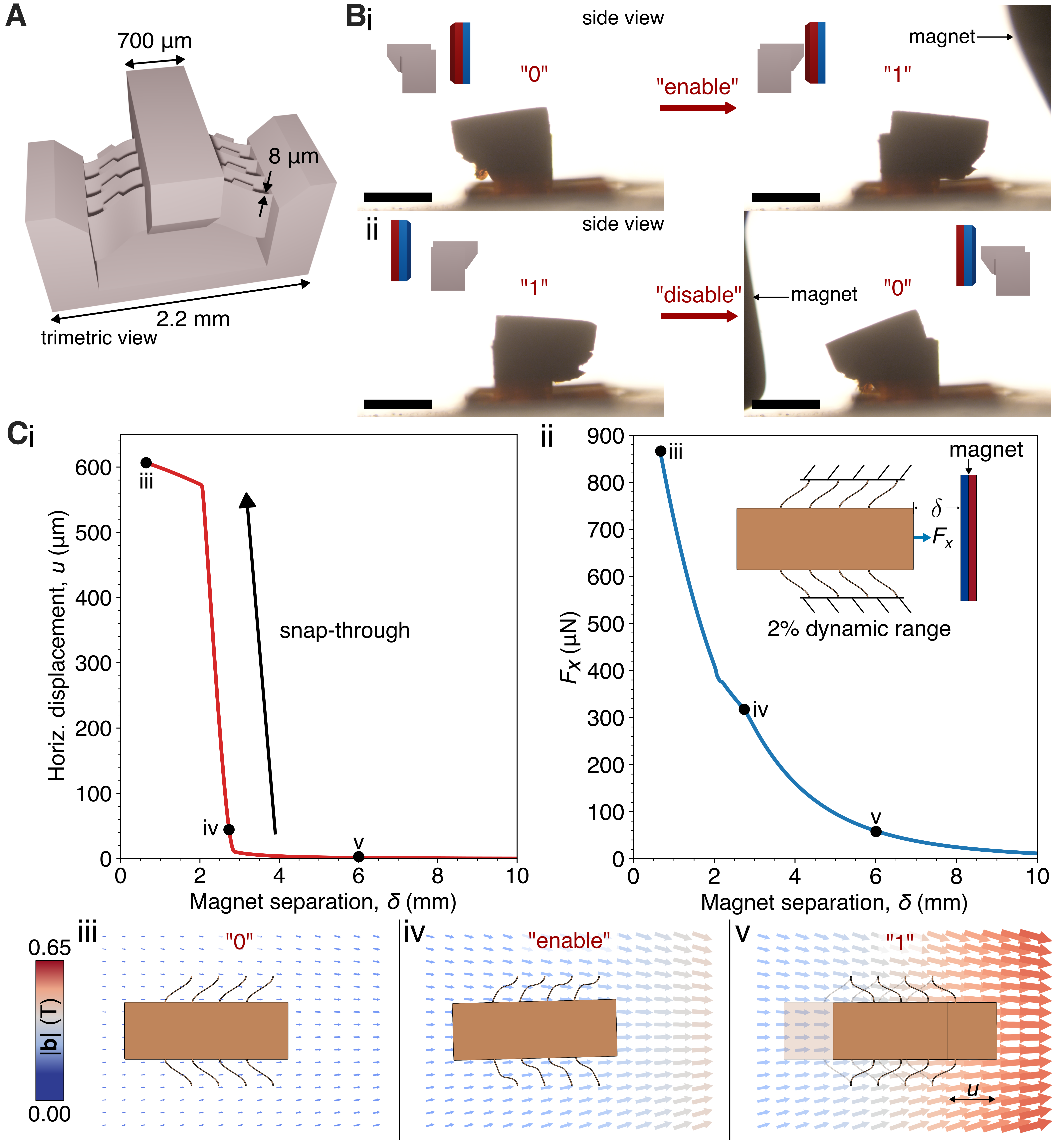}
\caption{\textbf{Functional \revision{bistable microstructure} via IONP nanocomposite.}
\revision{(A) Schematic of a bistable bit printed at 2\% dynamic range.}
(B) Experimental microscope images show a bistable ``bit" being (i) enabled from ``0" and ``1" and (ii) disabled from ``1" to ``0". All scale bars 1 mm. 
\revision{(C) Numerically calculated (i) horizontal displacement and (ii) magnetic force exerted on the surface of the central block on the 2\% dynamic range bistable bit are shown as a function of magnet separation $\delta$ (boundary conditions inset). The movement of the central block is shown progressing from states (iii) ``0", through (iv) ``enable", to (v) ``1" with increasing magnetic field intensity as the magnet approaches the bit.} 
\\\hspace{\textwidth} 
}
\label{fig:4applications}
\end{centering}
\end{figure}
}

\subsubsection{Bistable logic bit}
As a further demonstration of remote magnetic actuation for information storage, we fabricated \revision{and magnetically actuated a bistable IONP structure that functions as a logic bit (Figure 6A-B)}. \revision{We printed these structures onto an unmagnetized, solid polymer base, which provides optical transparency and mechanically anchors the bistable element to prevent rigid-body motion during actuation.} The bistable structure consisted of a homogeneous central block designed to move as a rigid body and a substrate with a series of thin, ribbon-like hinges \revision{around 8~\textmu{}m to 10~\textmu{}m thick (Figure S12)}. \revision{Computational modeling demonstrates that as the magnet separation $\delta$ from the central block is reduced to a critical value, the structure undergoes a snap-through transition between its bistable states (Figure 6C, panel i), corresponding with a limit-point instability in the force response (Figure 6C, panel ii), which occurs at a critical force of 315 \textmu{}N. Kinematically, this event coincides with local buckling of the hinges (Figure 6C, panels iii-v; Supplemental Video 5).} Hence, this bistable structure can function like a binary, non-volatile logic bit, static in either a ``0'' or a ``1'' position (Figure 6B; Supplemental Video 6). \revision{We numerically calculated the response time of the bistable bit to be ${\sim}100$ ms (Figure S13B).}

When a magnet was introduced in the proximity of a structure containing a sufficient concentration of IONPs, the magnetic force was sufficient to pull the block between the two stable configurations. By changing both the incident dose of the entire bit (i.e., varying the spatial distribution of magnetic nanoparticles) and the magnitude of the applied magnetic field gradient, we created a design map of actuatable configurations. Figure 7A shows the experimentally determined permissible design space (center). When the dose and field gradient magnitude are compatible, the magnet can be used to toggle the bit between stable configurations (Figure 7A, panels i-vi; Supplemental Videos 7, 8). Conversely, in the incompatible regime, the bit moves towards the magnet, but the magnetic stimulus does not provide enough force to overcome the bistable energy barrier (Figure 7A, panels vii-ix). While the stiffness of the ribbon-like hinges plays a role in the ability of the bit to overcome this energy barrier, this phase diagram demonstrates that the compatibility of each system is due to a coupling between mechanical and magnetic properties.

This result can be applied to a variety of potential applications in microscale devices. For example, Figure 7B depicts an array of bistable bits which encode a sequence of values (e.g., ``101'') via spatial patterning of polymerization. Namely, the middle bit is printed at a different dose (e.g., 100\% dynamic range) than the others (e.g., 36\% dynamic range). When the applied field gradient is too low (e.g., $\ll 0.03$ T mm\textsuperscript{-1}), none of the bits actuate. Conversely, when the applied field gradient is too strong (e.g., $\geq 0.12$ T mm\textsuperscript{-1}), all of the bits actuate. However, when the correct field strength (e.g., 0.08 T mm\textsuperscript{-1}) is applied, the bits selectively actuate according to the IONP distribution, and the encoded message is revealed. In this case, the magnet associated with the correct field strength acts like a required ``key'' which unlocks the desired response. To demonstrate this concept experimentally, we printed a set of bits using these differing doses, showing that a single magnet can selectively actuate only the correct bits (Figure 7B, panel ii; Supplemental Video 9). In practice, this concept can be extended to pattern a large array which encrypts a desired message, unlocked only by using the correct magnetic key.
\begin{figure}[h]
\begin{centering}
\includegraphics[width=0.85\textwidth]{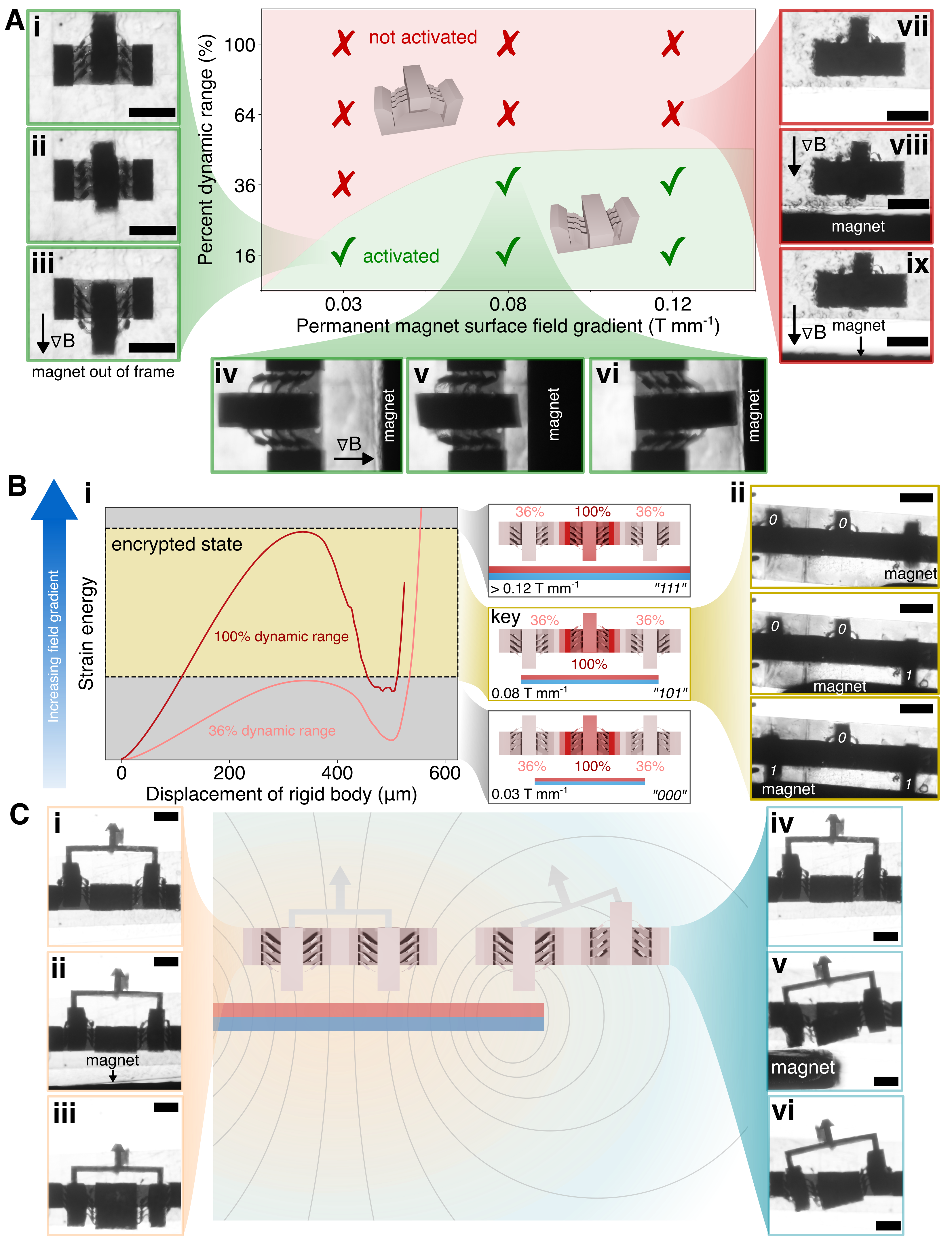}
\caption{\textbf{Non-volatile information in a magnetically bistable bit.} (A) Printing the bistable bit at different percentages of the dynamic range enables different states of activation (i.e., switching to the ``1" state) as a function of field gradient strength by using different magnets. (i-iii) The bit at 16\% dynamic range activates for all tested field gradients, demonstrating remotely-actuated snap-through. (iv-vi) The bit at 36\% dynamic range activates only for the second highest field gradient, requiring actuation at close proximity. (vii-ix) The bit at 64\% dynamic range moves towards the magnet but does not activate in the highest field gradient. (B) Encryption is enabled by printing bits at different doses, (i) requiring the correct magnetic field gradient as a ``key" to provide a specific amount of energy to access the encrypted state of ``101" in this case; (ii) this is experimentally demonstrated with the 0.08 T mm\textsuperscript{-1} field strength magnet. (C) An assembly of bistable bits is sensitive to the spatially varying field associated with a permanent magnet, behaving like a sensor near the (i-iii) center and (iv-vi) side of the magnet. All scale bars 1 mm. 
}
\label{fig:5}
\end{centering}
\end{figure}
Finally, we show how an assembly of these bistable bits is sensitive to the spatially varying field associated with a permanent (e.g., bar) magnet, yielding a structure that acts like a field gradient sensor (Figure 7C). We printed two identical bits in proximity, connected by a thin arrow-shaped indicator (the indicator does not affect the bistability of the bits). When the magnetic field and its gradient are symmetric about the centerline between two neighboring bits, they both actuate and the indicator remains straight (Figure 7C, panels i-iii). However, when the applied field is asymmetric, the bit assembly selectively actuates and the indicator is biased towards one side, forming a non-volatile indication of the direction of the stronger field gradient (Figure~ 7C, panels iv-vi). Using the phase map of compatible dose-field strength conditions developed previously, the sensitivity and spatial behavior of such an assembly can be readily tuned.

\section{Discussion}
We presented a fabrication route along with systematic characterization experiments to create programmable magnetic materials at the microscale. By harnessing an apparent ``core-shell effect" that affects nanoparticle distributions with varying hydrogel crosslink density, we fabricated materials with 3D microscale features and spatially tunable nanoparticle distribution. We demonstrated this effect through EDS characterization, showing iron oxide nanoparticle (IONP) distributions that were controlled by the incident two-photon dose and illustrating how the crosslink density affects the diffusion kinetics of the fabrication process. We also characterized \revision{and numerically modeled} the \revision{coupled} magnetic and mechanical properties of these materials to inform the \revision{design, fabrication, and response} of more complex magnetically active structures with microscale features, demonstrated with a microscale gripper and a bistable bit. By modulating laser power during printing, we control spatial nanoparticle content within the parts, imbuing them with additional functionality as exemplified by the microscale gripper with arms that respond differently to the same magnet and the bit register that only reveals its message in the presence of a specific magnetic field gradient. 

\revision{At the same time, the current fabrication and characterization approach imposes several limitations. Actuation dynamics, such as the snap-through mechanism in the bistable bit, remain difficult to experimentally characterize due to the need to maintain the hydrogel in a hydrated state, which necessitates \emph{in situ} high-speed microscopy together with precise spatial control of the applied external field gradient.}  

\revision{Moreover, fatigue resistance remains a potential avenue for improvement; we measured the fatigue life of the bistable bit to be 24 back-and-forth actuation cycles before failure by substrate delamination (Supplemental Video 10). We attribute this failure mechanism to the mismatch in swelling between the hydrogel-IONP composite and the polymer substrate, together with the development of shear forces at the interface. As a result, improving interfacial adhesion between disparate materials will be critical for improving the robustness of these devices, particularly those which will undergo cyclic actuation in service.}
\revision{Finally, the present hydrogel system is sensitive to dehydration and brittle when dry, which generates internal stresses leading to bending and cracking, particularly for structures of high aspect ratios dried at atmospheric pressure. We observed that critical point drying (CPD) in carbon dioxide suppresses drying-induced fracture (Figure S12), providing one viable route towards applicability of our IONP composites in dry environments.}

Despite these limitations, magnetically actuated hydrogels are still of great interest in medical applications due to their robust mechanical properties and functionality when used as remotely controlled microrobots and devices \cite{giltinan20213d, ceylan20183d, ceylan20213d, peters2014superparamagnetic}. \revision{We have summarized a variety of recent applications of soft-magnetic microscale composites in Table S3; the majority of these microdevices operate in or are compatible with an aqueous environment.} \revision{Precise remote actuation of microscale, magnetically-responsive structures has also been enabled by recent advances in electromagnetic coil-based systems, which can provide spatial control of field gradients and thus driving forces for soft-magnetic motion\cite{xing2025trimag, Yasa2019}, making real-world remote actuation of these structures realistically feasible.} Our fabrication approach \revision{complements these advances by enabling both precise control of microscale structure and direct spatial encoding of functionality during printing}, paving the way for tunable magnetically responsive microscale metamaterials, robots, and devices. 

\section{Methods}
\subsection{Formulation of PEGDA-based photosensitive hydrogel resin}
To synthesize the two-photon-sensitive hydrogel resin, we mixed two solutions: one containing poly(ethylene glycol) diacrylate (PEGDA, molecular weight 700; Sigma-Aldrich) with deionized water in equal parts by volume, and a second containing 6.2 mg of photoinitiator 7-diethylamino-3-thenoylcoumarin (DETC; Luxottica Exciton) dissolved in 465 \textmu{}L of dimethyl sulfoxide (DMSO; Sigma-Aldrich). We prepared the hydrogel resin for printing by mixing the DETC solution into the PEGDA solution at a volume ratio of 1:10, ultrasonicating for 60 s.

\subsection{3D-printing via TPP}\label{sec:printing}
We used the NanoOne two-photon polymerization printer (UpNano, Vienna, Austria) operating in vat mode to print structures using the hydrogel resin. Structures were printed onto clean borosilicate glass substrates at a linear scanning speed of 600 mm s$^{-1}$ with varying laser powers as described in the text and SI section A1. Slicing and hatching distances were 5 \textmu{}m and 4.2 \textmu{}m, respectively. After printing, structures were washed in deionized water for a minimum of 30 minutes.

\subsection{Coprecipitation process}\label{process_flow}
Important process parameters associated with the coprecipitation process are summarized in Table S1.

\paragraph{Iron infusion.} To create the iron solution for infusion, we dissolved 7.29 g iron (III) chloride hexahydrate (FeCl$_3 \cdot$6H$_2$O, Sigma-Aldrich) and 2.98 g iron (II) chloride tetrahydrate (FeCl$_2 \cdot$4H$_2$O, Sigma-Aldrich) in deionized water to a total of 10 mL, ultrasonicating until all powder was dissolved. This creates an aqueous solution of 1 M iron (II) and 1.8 M iron (III) chloride. We then immersed the printed precursors in the iron solution for 60 minutes at 65$\degree$C. Following the immersion, the structures were placed in 1-methoxy-2-propanol acetate (PGMEA, Sigma-Aldrich) for 5 minutes at room temperature to separate the remaining salt solution from the microstructures.
\paragraph{Nanoparticle growth.}
To grow the nanoparticles, we then immersed the parts for 5 minutes at 50$\degree$C in ammonium hydroxide (NH$_4$OH, Sigma-Aldrich), followed by deionized water for 60 minutes at room temperature as a final rinse.
\paragraph{Drying.}
Where indicated, all samples were dried in ambient conditions (room temperature 23$\degree$C and atmospheric pressure) for at least one hour after the nanoparticle growth step. 

\subsection{Energy dispersive X-ray spectroscopy (EDS)}\label{sec:eds}
\revision{EDS measurements were collected on a Zeiss Gemini 450 scanning electron microscope with a Oxford AZtec 100 EDS detector.} All scans were performed in a single pass with constant settings of a 500 ms pixel dwell time for line scans comprised of 500 points and a 100~\textmu{}m pixel dwell time for 1024 channel elemental maps. Line scans were smoothed with a Savitzky-Golay filter by fitting a third-order polynomial to every 45 data points. \revision{After removal from water, excess water was removed from around the cylinders, and the cylinders were immediately cut to maximize cylinder hydration and reduce cracking before cutting. More information on the sample preparation and data processing scheme can be found in Figure S1.}

\subsection{X-ray diffraction}
Samples were measured using a molybdenum X-ray source at 0.7093 \r{A} on the Malvern Panalytical Empyrean from angles $8\degree$ to $60\degree$ centered on sample of dimensions 5000 $\times$ 5000 $\times$ 500~\textmu{}m$^3$. 

\subsection{Magnetic actuation}
The different-sized NdFeB magnets used for varying permanent magnet surface field gradient in Figure 7 were procured from K\&J Magnetics, Inc. In particular, we used BX041-N52, BX042-N52, and BX044-N52 for field gradients 0.03 T mm\textsuperscript{-1}, 0.08 T mm\textsuperscript{-1}, and 0.12 T mm\textsuperscript{-1}, and obtained these field gradient values as published by the manufacturer. 

\subsection{Coupled magneto-elastic simulations}
Further details regarding coupled magneto-elasticity theory and the modeling approach can be found in SI sections A9 and A10, respectively. The modeled referential fluid-solid subdomains are depicted in Figure S14 and a representative mesh and boundary conditions are illustrated in Figure S15.

\clearpage

\section*{Resource availability}
\subsection*{Lead contact}
Requests for further information and resources should be directed to and will be fulfilled by the lead contact, Carlos M. Portela (cportela@mit.edu).

\subsection*{Materials availability}
This study did not generate new unique reagents. 

\subsection*{Data and code availability}
All data of this study are available within the article and the supplemental information or from the corresponding author on reasonable request. \revision{The source codes and mesh files for the FEniCSx simulations shown in this work are available from the following GitHub repository: 
\begin{itemize}
\item \url{https://github.com/ericstewart36/micro_softmagnetics}
\end{itemize}}

\section*{ACKNOWLEDGMENTS}
C.M.P. acknowledges partial support from the MIT MechE MathWorks Seed Fund and the MIT-Switzerland Lockheed Martin Seed Fund. D.W.Y. and Y.J. acknowledges support from the Swiss ETH domain Strategic Focus Area – Advanced Manufacturing program. R.M.S. and A.Y.C. acknowledge financial support from the National Science Foundation through the Graduate Research Fellowship Program. This material is based upon work supported by the National Science Foundation Graduate Research Fellowship Program under Grant No. 2141064. Any opinions, findings, and conclusions or recommendations expressed in this material are those of the authors and do not necessarily reflect the views of the National Science Foundation. The authors thank Dr. Lei Wu for assistance with bistability designs, and Professor Lallit Anand for discussions on modeling approaches. The authors also thank Tomas Grossmark and Caroline Ross for assistance with vibrating sample magnetometer measurements. This work was performed in part in the MIT.nano Fabrication and Characterization Facilities. 

\section*{AUTHOR CONTRIBUTIONS}

Conceptualization, all authors; methodology, Y.J., \revision{E.M.S.}, D.W.Y.; investigation, R.M.S., A.Y.C., \revision{E.M.S.}; \revision{software, E.M.S.}; writing---original draft, R.M.S., A.Y.C.; writing---review \& editing, all authors; funding acquisition, D.W.Y., C.M.P.; supervision, D.W.Y., C.M.P.

\section*{DECLARATION OF INTERESTS}
The authors have filed patents on the \emph{in situ} precipitation technology. 

\clearpage

\setcounter{section}{0}
\renewcommand{\thesubsection}{A\arabic{subsection}}
\setcounter{subsection}{0}
\setcounter{figure}{0}
\renewcommand{\thefigure}{\textbf{S\arabic{figure}}}
\renewcommand{\thetable}{\textbf{S\arabic{table}}}

\section*{Supplemental Notes}

\subsection{Process parameters and resulting physical properties}
The following table summarizes the important process parameters associated with the co-precipitation process. A detailed process flow is given in Section 4.3 of the main text.

\begin{table}[h]
\begin{centering}

\begin{tabular}{|l|l|}
\hline
Iron (II) chloride concentration  & 1 M    \\ \hline
Iron (III) chloride concentration & 1.8 M  \\ \hline
Ion infusion in iron salt solution & 60 min, 65$\degree$C \\ \hline
Precipitation in ammonium hydroxide & 5 min, 50$\degree$C  \\ \hline
\end{tabular}
\caption{\textbf{Process parameters associated with the co-precipitation process.}}
\end{centering}
\end{table} 

The following table summarizes measured physical properties as a function of two-photon dose; see also Figure 3 and Figure~\ref{fig:SI-MASS}.
\begin{table}[h]
\begin{centering}
\begin{tabular}{|p{3cm}|p{4cm}|p{4cm}|p{3cm}|}
\hline
\textbf{Dose (pct. dyn. range)} & \textbf{Elastic modulus (MPa)} & \textbf{Saturation magnetization (MA/m)} & \textbf{Mass increase (percent)} \\ \hline
2 & $19.9 \pm 1.8$  & $0.00580 \pm 0.00083$                          & $32.2 \pm 3.8$                 \\ \hline
18 & $36.0 \pm 3.5$  & $0.00240 \pm 0.00183$                          & $19.0 \pm 9.3$                 \\ \hline
51  & $53.8 \pm 7.8$  & $0.00146 \pm 0.00067$                          & $16.2 \pm 5.3$                 \\ \hline
100  & $53.2 \pm 8.1$  & $0.00077 \pm 0.00022$                          & $9.8 \pm 4.2$                  \\ \hline
\end{tabular}
\caption{\textbf{Measured physical properties as a function of two-photon dose.}}
\end{centering}
\end{table} 

The following table contextualizes our fabrication process with existing methods for realizing soft-magnetic composite materials with microscale feature sizes.
\begin{table}[h]
\label{tab:comparison}
    \centering
    {\small
    \begin{tabular}{|p{1cm}|p{1.6cm}|p{1.8cm}|p{2cm}|p{1.7cm}|p{1.7cm}|p{2.5cm}|p{1.2cm}|}
        \hline
        \textbf{Ref.}
            & \textbf{Magnetic material}
            & \textbf{Matrix material}
            & \textbf{Fabrication method}
            & \textbf{Magnetic loading}
            & \textbf{Feature resolution (\textmu{}m)}
            & \textbf{Application} 
            & \textbf{Spatial tunability} \\
        \hline
        \textbf{This study}
            & \textbf{Fe$_3$O$_4$}
            & \textbf{PEGDA}
            & \textbf{\emph{in situ} nanoparticle growth}
            & \textbf{24.35 wt\% (2\% crosslink density)}
            & \textbf{8}
            & \textbf{Sensing, actuating, information encoding} 
            & \textbf{Yes} \\
        \hline
        \cite{Sivudu2009}
            & Fe$_3$O$_4$
            & Acrylamide
            & \emph{in situ} nanoparticle growth
            & 6 wt\%
            & unreported
            & Bioseparation, tissue engineering
            & No \\
        \hline
        \cite{Kim2011}
            & Fe$_3$O$_4$
            & PEGDA
            & UV photopatterning
            & unreported
            & 13.68
            & Microactuators with programmable magnetization axis & No \\
        \hline
        \cite{peters2014superparamagnetic}
            & Fe$_3$O$_4$
            & SU8 (photoresist)
            & TPP
            & 2 vol\%
            & 9
            & Swimming microrobots 
            & No \\
        \hline
        \cite{ceylan20183d}
            & Fe$_3$O$_4$
            & Gelatin methacryloyl
            & TPP
            & 6 mg/mL
            & 3.3
            & Swimming microrobots 
            & No \\
        \hline
        \cite{Yasa2019}
            & Fe$_3$O$_4$
            & TMPTA
            & TPP
            & 20 mg/mL
            & 3.33
            & Cell delivery 
            & No \\
        \hline
        \cite{cabanach2020zwitterionic}
            & Fe$_3$O$_4$
            & PEG
            & TPP
            & 12.5 mg/mL
            & 2
            & Non-immunogenic microrobots 
            & No \\
        \hline
        \cite{ceylan20213d}
            & Fe$_3$O$_4$
            & PEG
            & TPP
            & 1 mg/mL
            & 6.7
            & Swimming microrobots and microrollers 
            & No \\
        \hline
        \cite{cestarollo2022nanoparticle}
            & Fe powder
            & Ecoflex
            & Molding
            & 6 vol\%
            & 200
            & Haptic displays 
            & No \\
        \hline
        \cite{geid2024gym}
            & Fe$_3$O$_4$
            & Copolymer bilayer
            & CHiC crosslinking via fs laser
            & 11 wt\%
            & 5
            & Single-cell actuation 
            & No \\
        \hline
        \cite{xing2025trimag}
            & Fe$_3$O$_4$
            & PEGDA/ PETA
            & \emph{in situ} nanoparticle growth
            & unreported
            & 20
            & Microrobots for biological imaging and tracking 
            & No \\
        \hline
    \end{tabular}
    \caption{\revision{\textbf{Soft-magnetic composite materials at the microscale. }\textit{Abbreviations: }PEGDA, polyethylene glycol diacrylate; TPP, two-photon polymerization; TMPTA, trimethylolpropane triacrylate; PEG, polyethylene glycol; PETA, pentaerythritol tetraacrylate.}}}
\end{table}

\clearpage
\subsection{Converting laser power to dynamic range}
To ensure comparison across systems and variability in laser power across TPP printers or experimental campaigns, we present the laser power as a percent of dynamic range. This conversion was performed by scaling the dynamic range percentages quadratically with laser power as the dose $D$ is quadratically related to incident laser power $I$ and scan speed $v$, which remained constant for all prints in this work. Dynamic range $R$ for results shown correspond to the minimum incident laser power $I_{\text{min}}$ and fitting parameter $a$ in Equation (2) of the main text for the figures listed in Table \textbf{S4}. 

\begin{table}[hbt!]
\label{tab:dynRangeFitParams}
\begin{centering}
\begin{tabular}{|c|c|c|}
\hline
Main text figure  & $a$ (mW$^{-2}$)     & $I_{\text{min}}$ (mW) \\ \hline
1, 3-6 & $8.163 \times 10^{-4}$ & 150 \\ \hline
2      & $1.041 \times 10^{-3}$ & 165 \\ \hline
7      & $1.600 \times 10^{-3}$ & 200 \\ \hline
\end{tabular}
\caption{\textbf{Minimum incident laser power and fitting parameters used to calculate dynamic range $R$ for each main text figure.}}
\end{centering}

\end{table}

\clearpage
\subsection{Detailed EDS data}
Specimens were prepared for EDS measurements by slicing using a razor blade through each cross-section.

\begin{figure}[hbt!]
\begin{centering}
\includegraphics[width=0.9\textwidth]{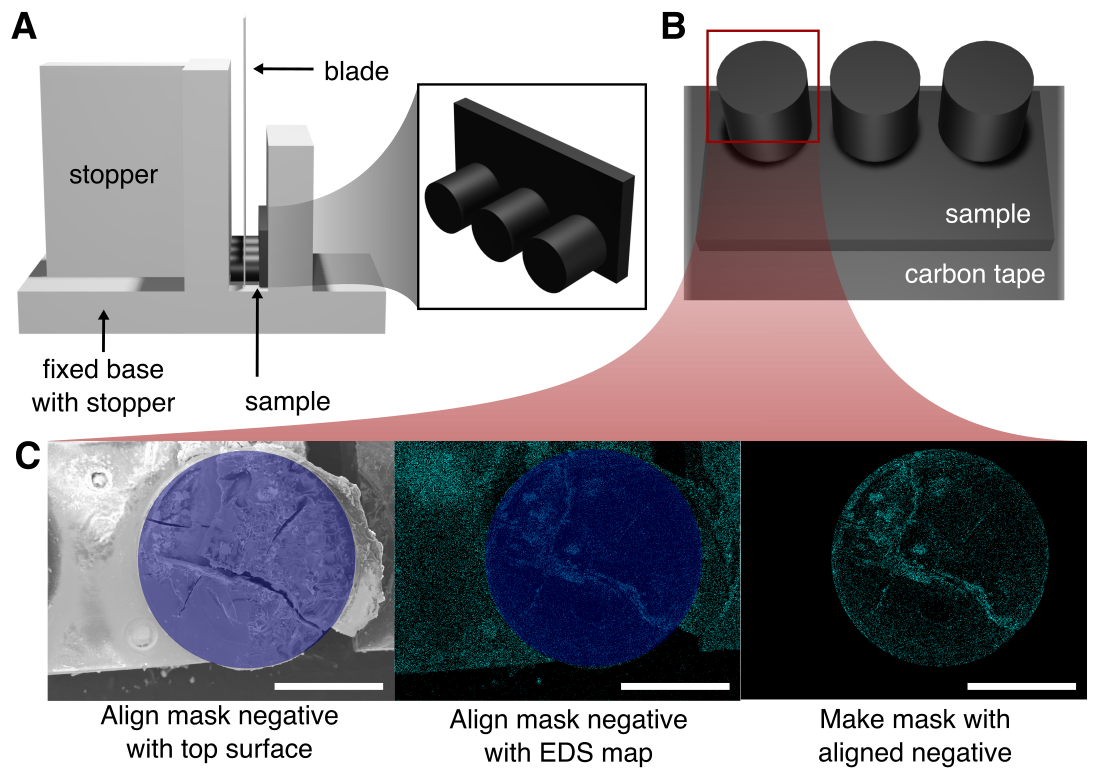}
\caption{\revision{\textbf{Procedure to cut, image, and mask EDS data.} (A) A set of cylinders is printed on a baseplate such that the cylinder sides are flat on a fixed base plate, then held in place with a stopper to cut cylinders with a uniform cross section for imaging. (B) Cut samples are mounted on carbon tape then imaged using a scanning electron microscope with a EDS detector. (C) Since the baseplate of the sample is comprised of iron, we mask the background from all images to improve visual clarity of the final image by aligning an elliptical mask negative with the desired cross section, then aligning the same negative with the corresponding EDS map relative to the edges of the image, and finally making a background mask with the aligned negative. All scale bars 500~\textmu{}m.}}
\label{fig:SI-EDSProcess}
\end{centering}
\end{figure}

\begin{figure}[hbt!]
\begin{centering}
\includegraphics[width=0.7\textwidth]{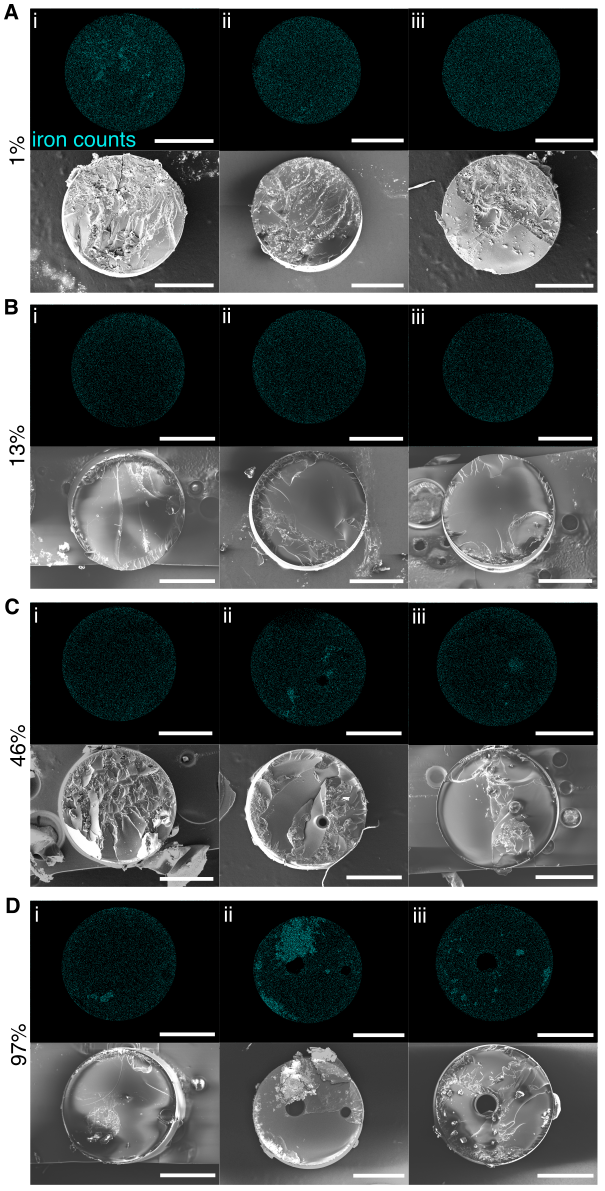}
\caption{\revision{\textbf{EDS data of cylinders of varying doses after ion infusion.} Cylinders of dynamic range percents of (A) 1\%, (B) 13\%, (C) 46\%, and (D) 98\% were infused with iron ions, then cut, dried, then imaged. (i-iii) Three separate samples per dose were imaged, demonstrating full infusion of iron ions into the center of the cylinder. All scale bars 500~\textmu{}m.}}
\label{fig:SI-InfusionSweep}
\end{centering}
\end{figure}

\begin{figure}[hbt!]
\begin{centering}
\includegraphics[width=0.95\textwidth]{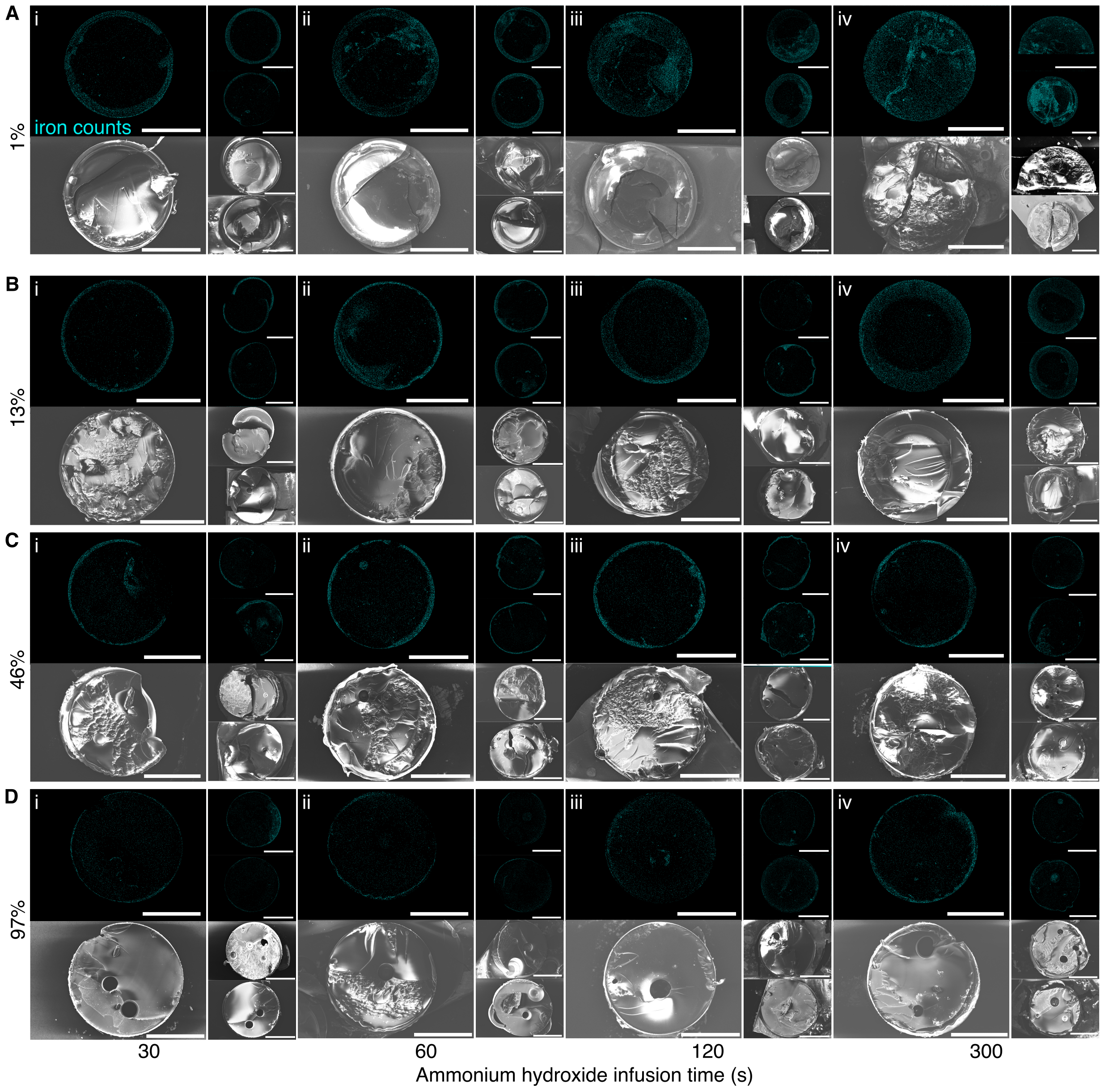}
\caption{\revision{\textbf{EDS data of cylinders after co-precipitation of varying durations.} Cylinders of dynamic range percents of (A) 1\%, (B) 13\%, (C) 46\%, and (D) 98\% were infused in ammonium hydroxide for durations of (i) 30 seconds, (ii) 60 seconds, (iii) 120 seconds, and (iv) 300 seconds, demonstrating varying degrees of a polymer-dominant core surrounded by a shell dominated by iron oxide. All scale bars 500~\textmu{}m. }}
\label{fig:SI-AmmoniumSweep}
\end{centering}
\end{figure}

\begin{figure}[hbt!]
\begin{centering}
\includegraphics[width=0.7\textwidth]{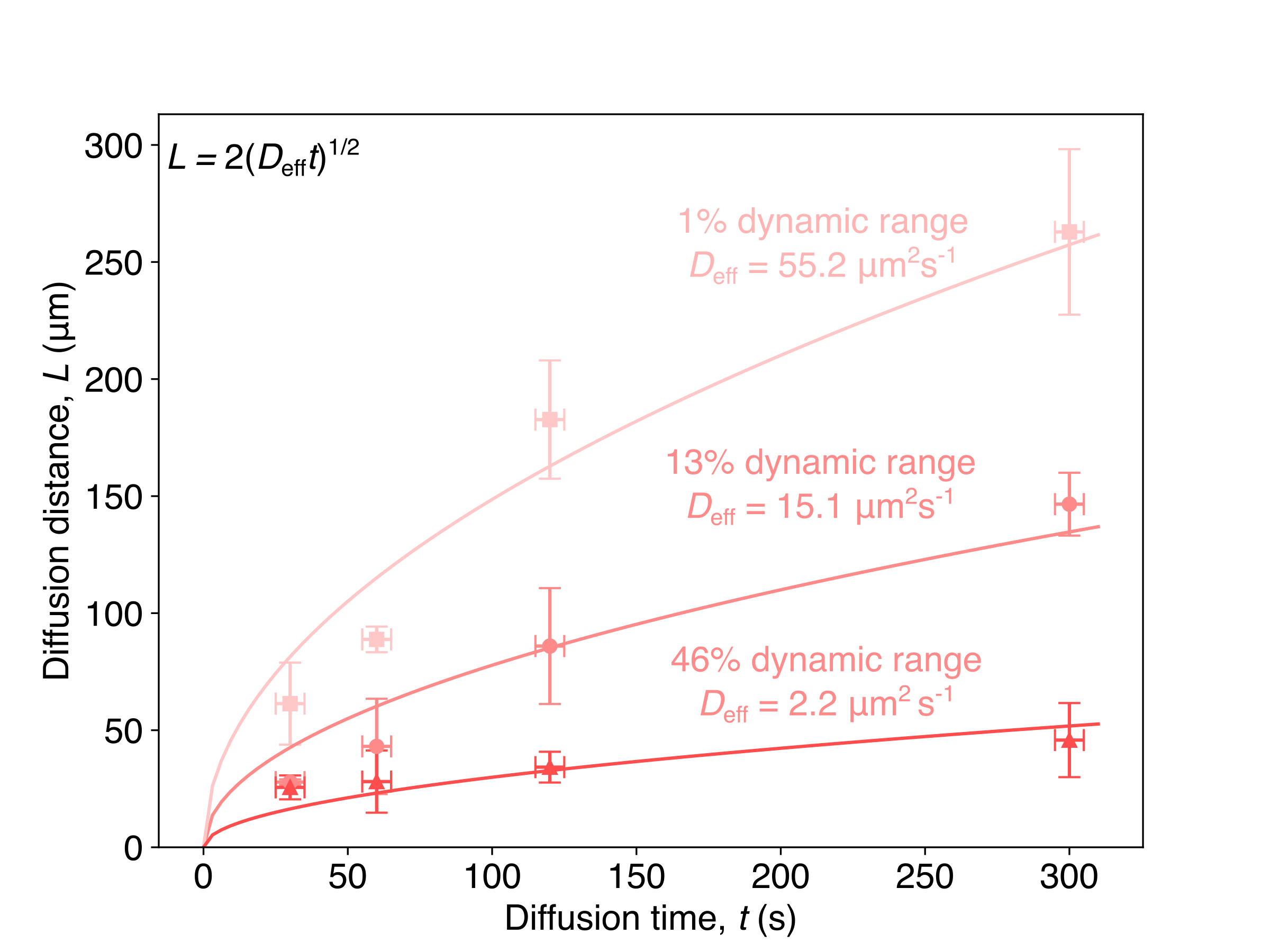}
\caption{\revision{\textbf{Diffusion distance as a function of ammonium hydroxide infusion time.} Diffusion distance $L$ was measured as an average of 5 measurements of shell thicknesses across different radial locations of 1\%, 13\%, and 46\% dynamic range 1 mm diameter cylinders from Figure~\ref{fig:SI-AmmoniumSweep}A-C and plotted as a function of diffusion time $t$, assuming a 5 second standard deviation of time of extraction from ammonium hydroxide. A best-fit curve is used to determine the effective diffusion coefficient, $D_\mathrm{eff}$}.}
\label{fig:SI-DiffusionDist}
\end{centering}
\end{figure}

\begin{figure}[hbt!]
\begin{centering}
\includegraphics[width=0.9\textwidth]{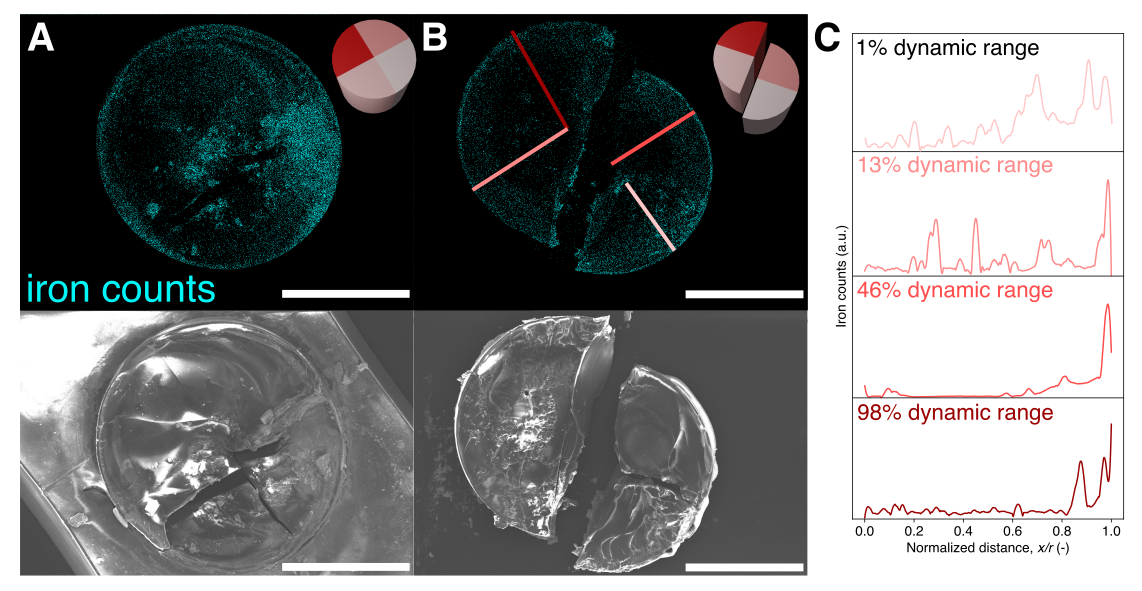}
\caption{\revision{\textbf{EDS data of cylinders printed with varying doses across a single cross section.} Cylinders with varying dynamic range percents of 1\% (light pink), 13\% (pink), 46\% (dark pink), and 98\% (red) were printed with four doses across the circular cross section of the cylinder, then coprecipitated and imaged. (A) a single sample on a baseplate is shown with varying intensities across the four different quadrants. (B) A cylindrical sample that cracked during drying is shown, with (C) line scans corresponding to varying intensities of iron counts across four cylinder quadrants. All scale bars 500~\textmu{}m.}}
\label{fig:SI-MultiLP}
\end{centering}
\end{figure}

\begin{figure}[hbt!]
\begin{centering}
\includegraphics[width=0.9\textwidth]{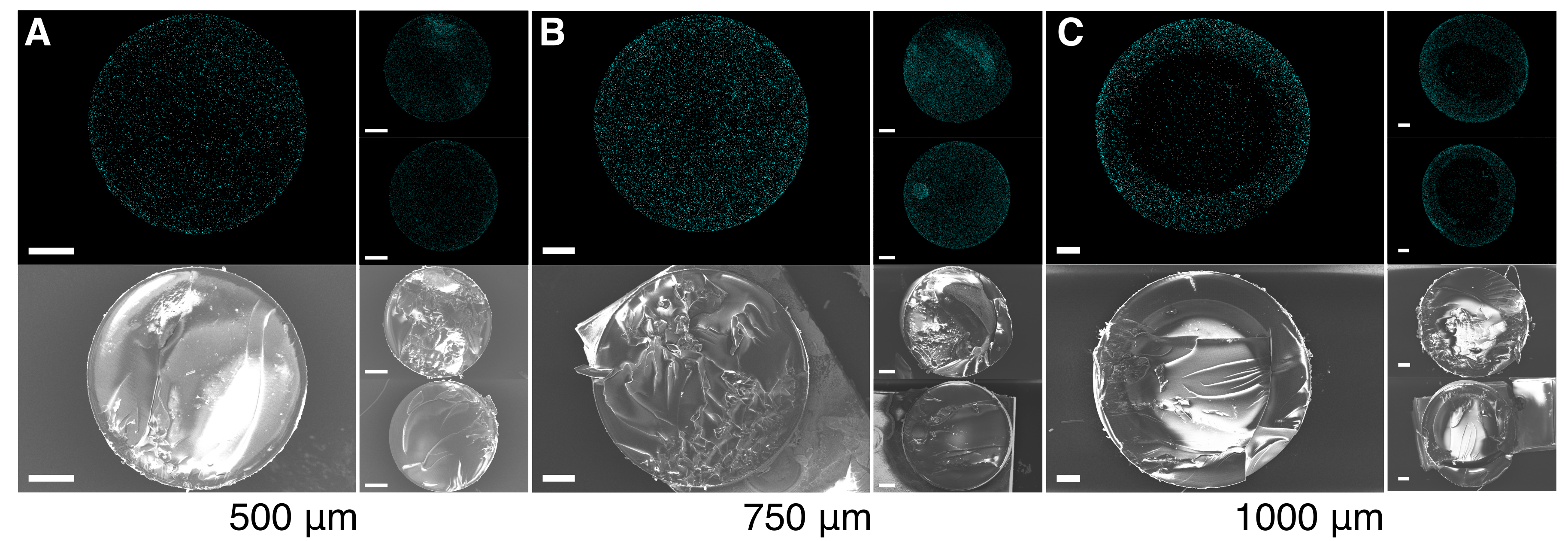}
\caption{\revision{\textbf{EDS data of cylinders printed with varying diameters.} Cylinders of constant 13\% dynamic range at varying diameters from (A) 500~\textmu{}m to (B) 750~\textmu{}m to (C) 1000~\textmu{}m were coprecipitated and imaged, demonstrating uniform nanoparticle distribution for sizes ~750~\textmu{}m and smaller. All scale bars 100~\textmu{}m.}}
\label{fig:SI-cylinderSize}
\end{centering}
\end{figure}

\clearpage
\revision{\subsection{Further physical characterization}}
\begin{figure}[hbt!]
\begin{centering}
\includegraphics[width=0.9\textwidth]{./figures/SFIG_HydratedProps.png}
\caption{\revision{\textbf{Magnetic and mechanical properties of the hydrated IONP composite material.}
(A) Magnetization curves of monolithic 5 $\times$ 5 $\times$ 0.5 mm\textsuperscript{3} blocks printed with different laser powers: 1\% (light pink), 13\%, 46\%, and 98\% (red) dynamic range. Shaded regions indicate one standard deviation from the mean (solid lines) across three sample replicates. 
(B) Stiffness (red triangles) are plotted against percent of dynamic range. An example stress-strain curve with stiffness obtained from the linear-loading-regime slope (red line) are shown in the inset. }}
\label{fig:SI-HydratedProps}
\end{centering}
\end{figure}
\clearpage

\begin{figure}[hbt!]
\begin{centering}
\includegraphics[width=0.7\textwidth]{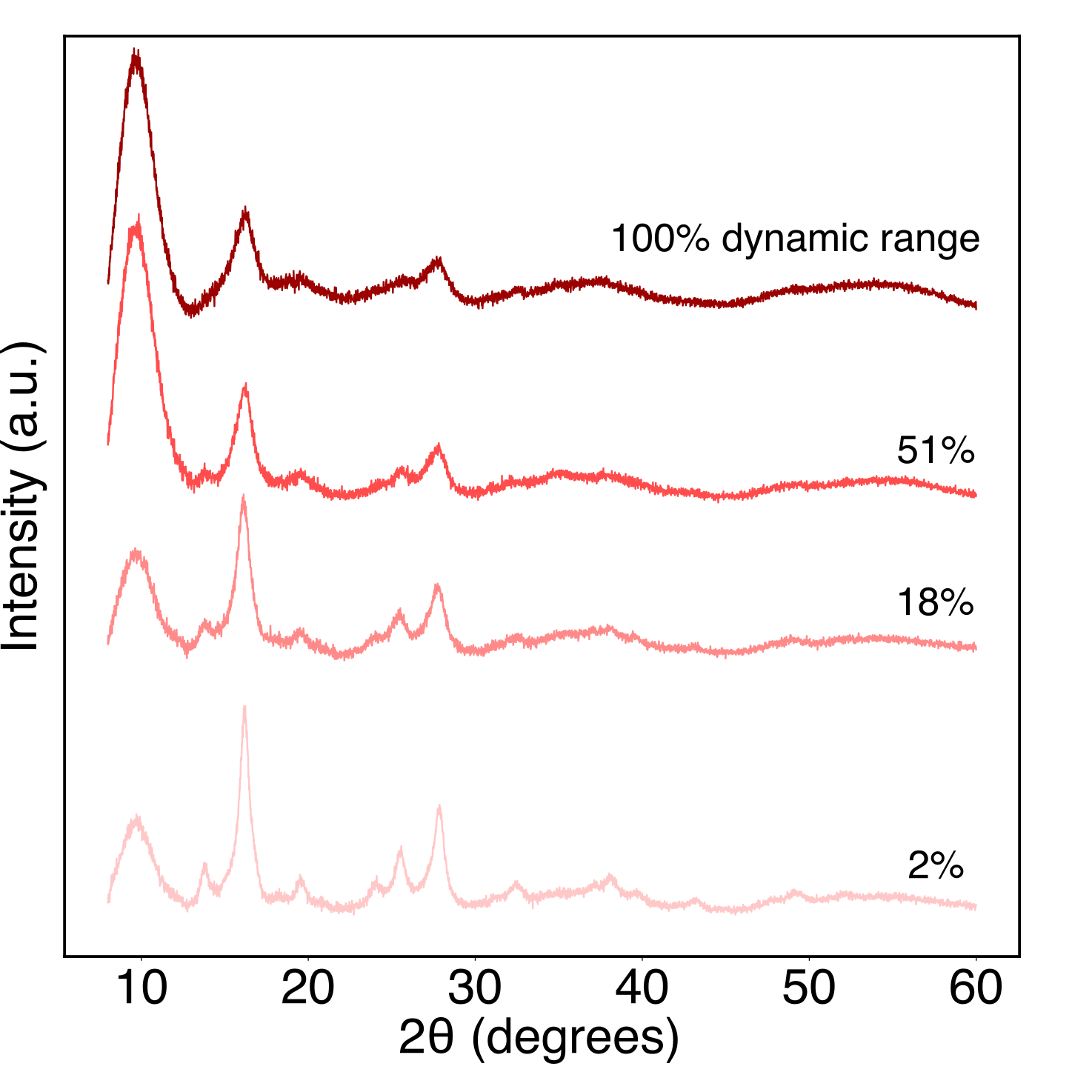}
\caption{\textbf{X-ray diffraction demonstrates a signal consistent with the presence of IONPs in the composite.} }
\label{fig:SI-XRD}
\end{centering}
\end{figure}
\clearpage

\begin{figure}[hbt!]
\begin{centering}
\includegraphics[width=0.7\textwidth]{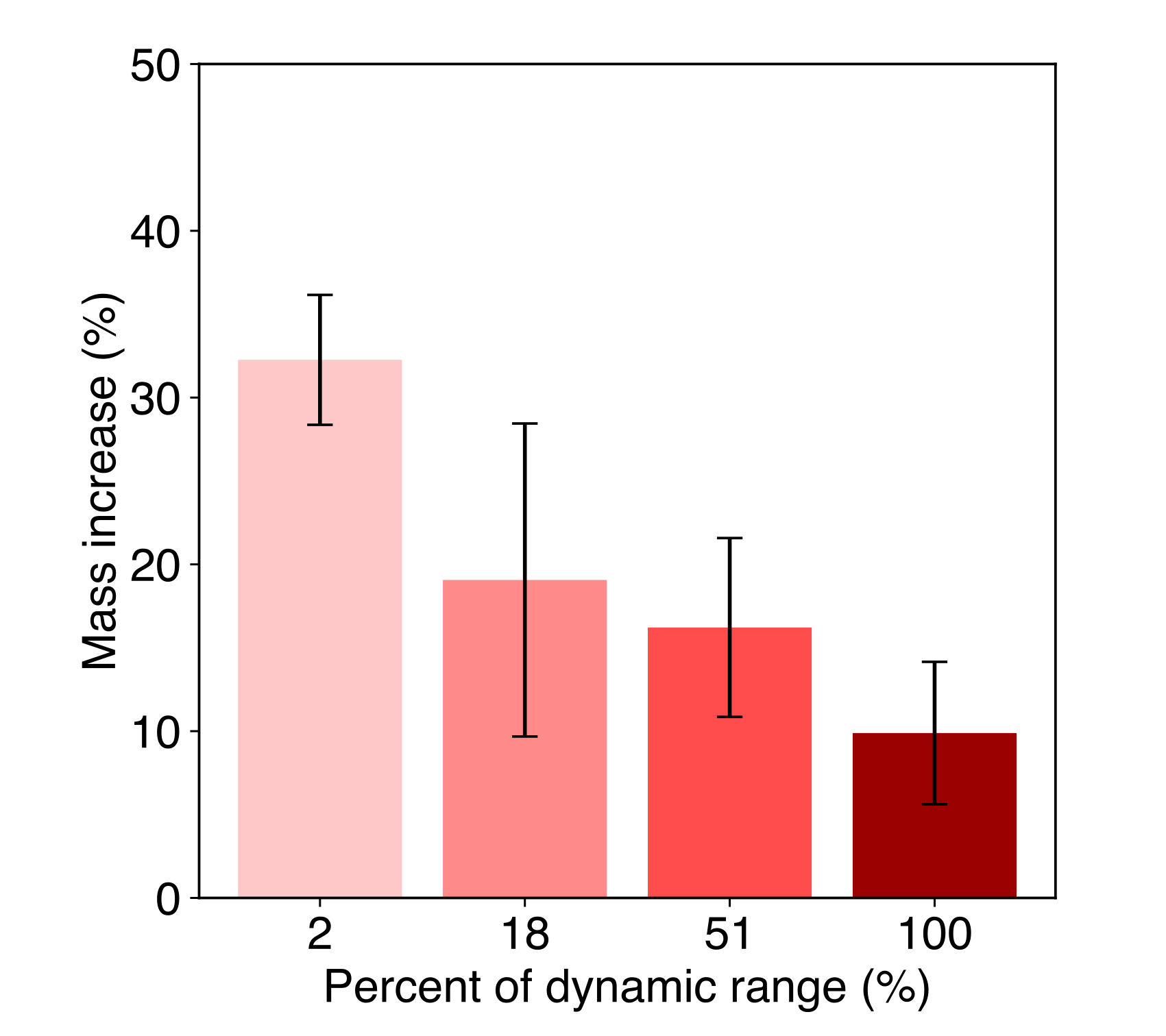}
\caption{\textbf{Mass change after co-precipitation as a function of percent of dynamic range (two-photon dose), suggesting that as the two-photon dose increases, the core-shell effect dominates and the average IONP concentration decreases.} }
\label{fig:SI-MASS}
\end{centering}
\end{figure}

\clearpage
\subsection{Compliance correction}
To facilitate measuring the mechanical properties of an array of individual monolithic pillar specimens, we printed and coprecipitated the pillar array on a large, flat monolithic baseplate printed at 400 mW. The quasi-static mechanical data (stiffness and yield strength) reported in the main text correspond to the properties of the pillars alone, after the following baseplate compliance correction procedure was undertaken.

During the experiment, we measure an effective elastic modulus $E_\mathrm{eff}$ which is the series equivalent of the pillar elastic modulus $E_1$ and the baseplate elastic modulus $E_2$, i.e., 

$$E_\mathrm{eff} = \left(\frac{1}{E_1} + \frac{1}{E_2}\right)^{-1}.$$

Here, because of the large, flat nature of the baseplate, $E_2$ represents a combined property of the material and geometry (i.e., not a true material property, as in the case of $E_1$). To measure $E_2$ directly, we perform a second set of experiments in which the indenter is used to compress the baseplate by a small amount relative to its diameter. For a circular-cylindrical flat punch of radius $a$ indenting an isotropic linear elastic half-space with material properties $E_2$ and $\nu$, Harding and Sneddon \cite{Harding1945} report the load-displacement relation

$$P = \frac{2E_2a}{1-\nu^2} \delta.$$

The baseplate modulus $E_2$ is extracted from this equation together with the measured values of $P$ and $\delta$ together with known $\nu = 0.35$ for printed laser power 400 mW, and a tip size $a = 200$ \textmu{}m. Across three experiments, we compute $E_2 = 67.5 \pm 0.7$ MPa. Finally, for each laser power, we compute
$$E_1 = \left(\frac{1}{E_\mathrm{eff}} - \frac{1}{E_2}\right)^{-1},$$
which is reported in Fig. 3B of the main text.

\clearpage
\subsection{Stress relaxation experiments}
Although the mechanical compression experiments were performed quasi-statically (at a nominal strain rate of 10\textsuperscript{-3} s\textsuperscript{-1}), the structure of the crosslinked polymer imbues the IONP composite with rate dependent properties. To understand the extent of this rate dependence, we conducted stress-relaxation experiments on monolithic, rectangular pillars of the same nominal dimensions as the ones used in the uniaxial compression experiments. We model the rate-sensitive material using a 4-branch Prony series, with relaxation modulus

$$E_r(t) = E_0 + \sum_{i=1}^4 E_i \exp\left(\frac{-t}{\tau_i}\right),$$

where $E_0$, $E_1$, $\dots$, $E_4$, $\tau_1$, $\tau_2$, $\dots$, $\tau_4$ are material parameters.

Using the Alemnis ASA nanoindenter, we applied a ramped input strain profile followed by a hold at a constant strain. This prescribed input strain profile differs from an ideal stress relaxation experiment due to the presence of a finite strain rate during loading. To accurately capture the effects of this strain input in the resulting stress output, we modeled the input profile as $$\varepsilon(t) = kt - k(t-t_1)[h(t-t_1)],$$ where $k$ and $t_1$ are input parameters and $h(\cdot)$ represents the Heaviside step function. Using least-squares regression, we fit the experimental data to the resulting output stress-time profile associated with the 4-branch Prony series, $$\sigma(t) = kt_1 E_0 + \sum_{i=1}^4 \tau_i k E_i \left[\exp\left(\frac{-(t - t_i)}{\tau_i}\right) - \exp\left(\frac{-t}{\tau_i}\right)\right].$$

\begin{figure}[hbt!]
\begin{centering}
\includegraphics[width=0.6\textwidth]{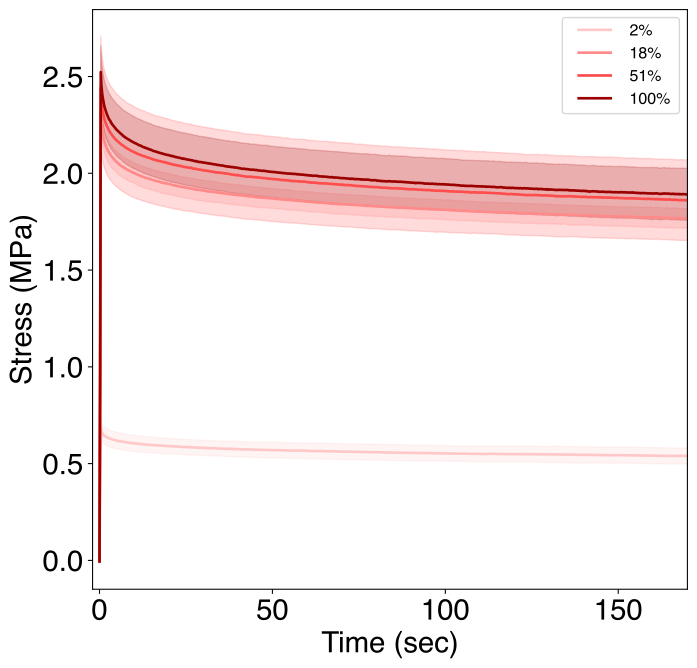}
\caption{\textbf{Stress-relaxation data for different dynamic range percents.} Dose fitting parameters correspond to $a = 8.163 \times 10^{-4}$ mW$^{-2}$ and $I_{\text{min}} = 150$ mW. }
\label{fig:SI-relaxation}
\end{centering}
\end{figure}

The 4-branch model was identified as the minimum number of branches for which the least-squares error stabilized. Figure~\ref{fig:SI-relaxation} shows the mean (solid line) and a $\pm 1$ standard-deviation around this mean (shaded regions) response for varying incident laser power.

\clearpage
\subsection{Measurement of the Poisson's ratio}
\begin{figure}[hbt!]
\begin{centering}
\includegraphics[width=0.7\textwidth]{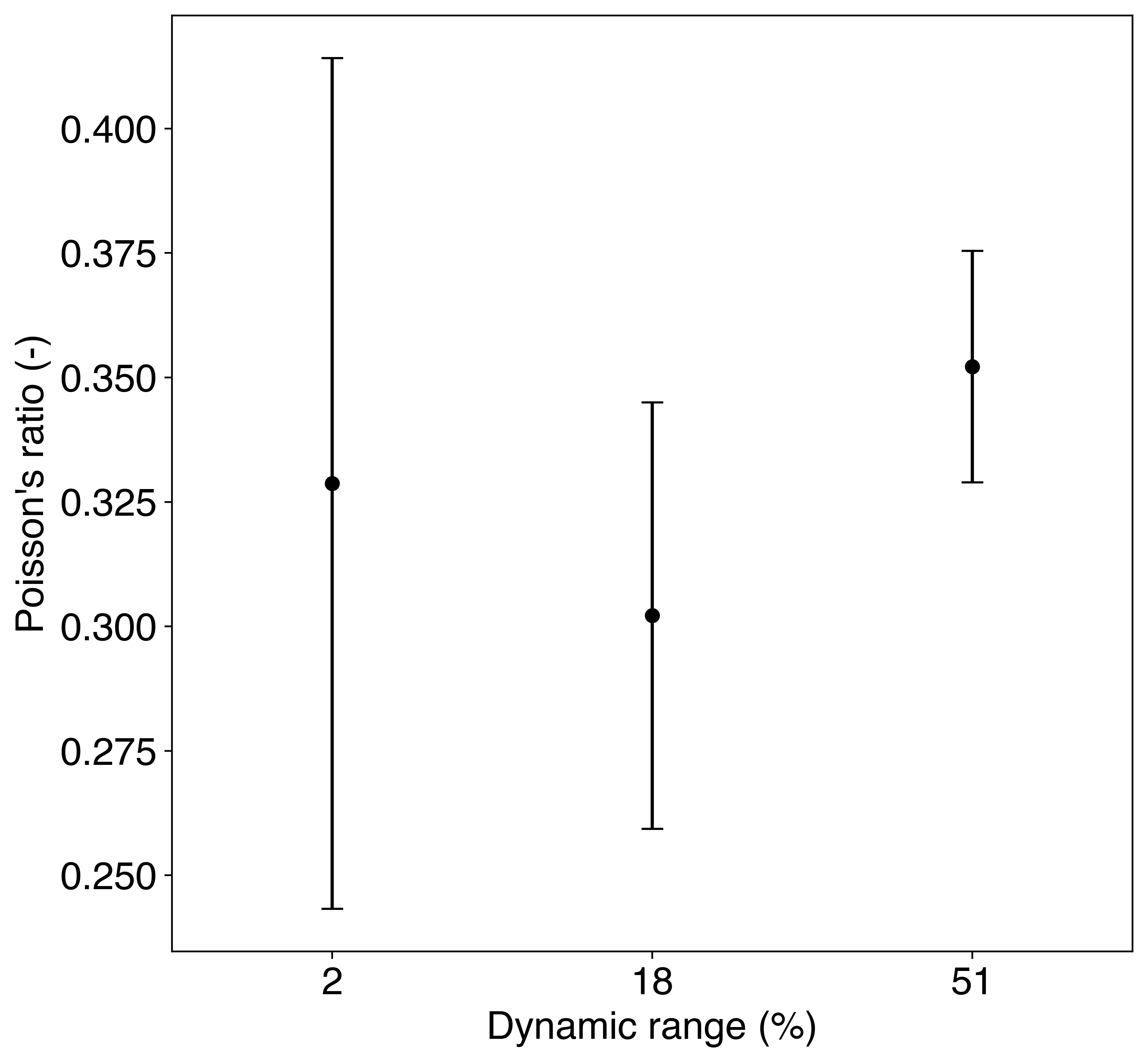}
\caption{\textbf{Poisson's ratio of the IONP composites as a function of printing laser power.} Dose fitting parameters correspond to $a = 8.163 \times 10^{-4}$ mW$^{-2}$ and $I_{\text{min}} = 150$ mW.}
\label{fig:SI-poisson}
\end{centering}
\end{figure}

In a state of uniaxial stress, the Poisson's ratio $\nu$ of an isotropic linear elastic material is defined as 
$$\nu \equiv -\frac{\varepsilon_{\mathrm{lat}}}{\varepsilon_{\mathrm{long}}},$$

where $\varepsilon_{\mathrm{long}}$ is the strain in the direction parallel to the applied stress, and $\varepsilon_{\mathrm{lat}}$ is the strain transverse to the applied stress direction. To determine the effective Poisson ratio of the IONP composites as a function of laser power (Fig.~\ref{fig:SI-poisson}), we tracked the axial and lateral strains during the initial, linear portion of uniaxial compression experiments done \textit{in situ} within a scanning electron microscope. Point tracking was performed in Blender by tracking lateral and axial extreme edge points and extracting their coordinates to calculate change in width and height of the rectangular pillars.

\clearpage
\subsection{\revision{Bistable bit geometry and response time}}
\revision{The bistable bit provides a representative example of the minimum achievable characteristic length scale of this fabrication process. Because of IONP agglomeration on the surface, the coprecipitated sample appears somewhat larger than the as-printed sample. }
\begin{figure}[hbt!]
\begin{centering}
\includegraphics[width=0.7\textwidth]{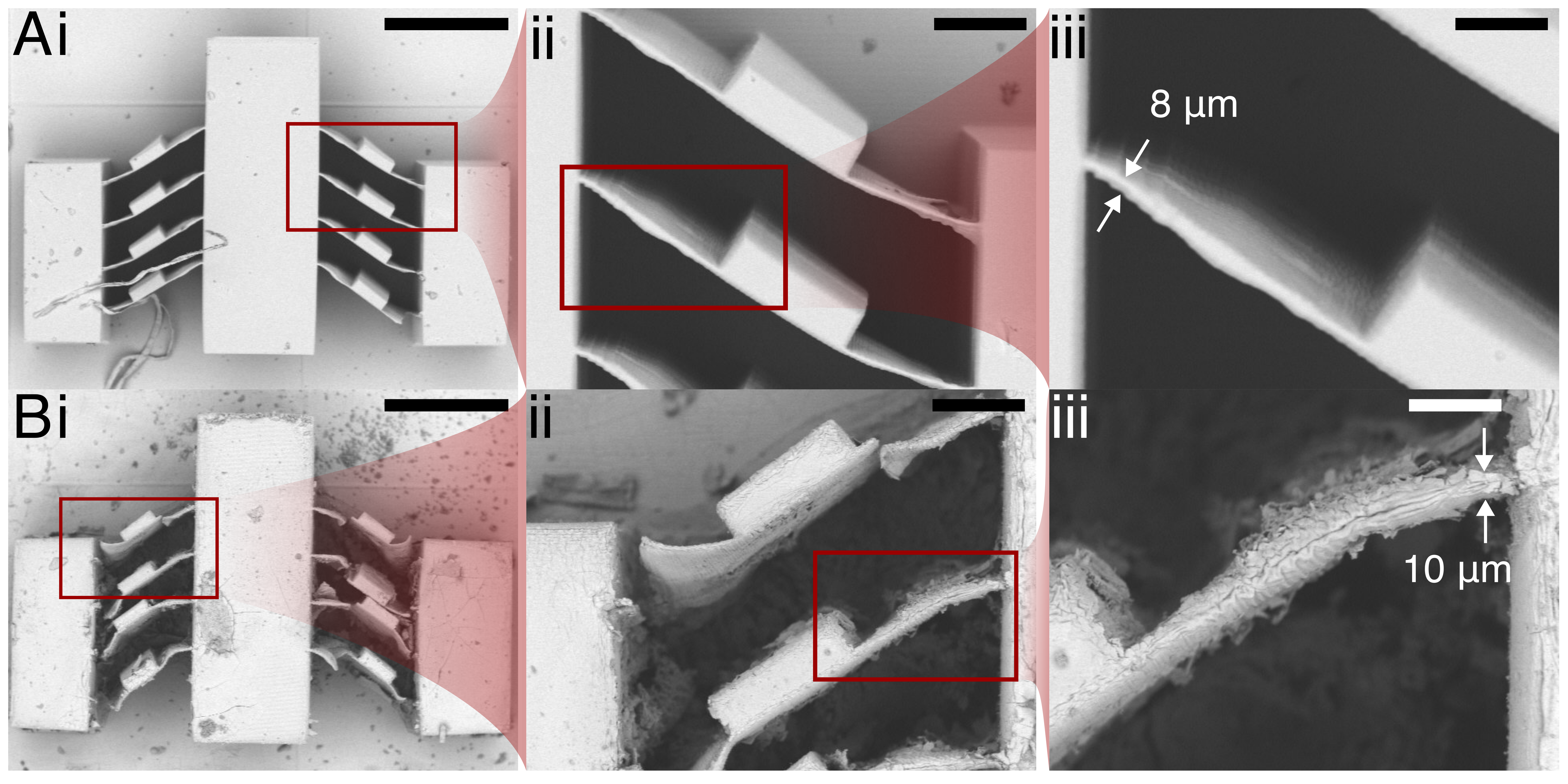}
\caption{\revision{\textbf{Scanning electron microscope images of dried bistable bits.} (A) As-printed and (B) coprecipitated bistable samples processed with critical point drying are shown with magnifications spanning (i) the whole sample, (ii) two bistable beams, and (iii) a single bistable beam with annotated thicknesses. Scale bars (i) 500~\textmu{}m, (ii) 100~\textmu{}m, and (iii) 50~\textmu{}m.} }
\label{fig:SI-charLengthScale}
\end{centering}
\end{figure}

\begin{figure}[hbt!]
\begin{centering}
\includegraphics[width=0.7\textwidth]{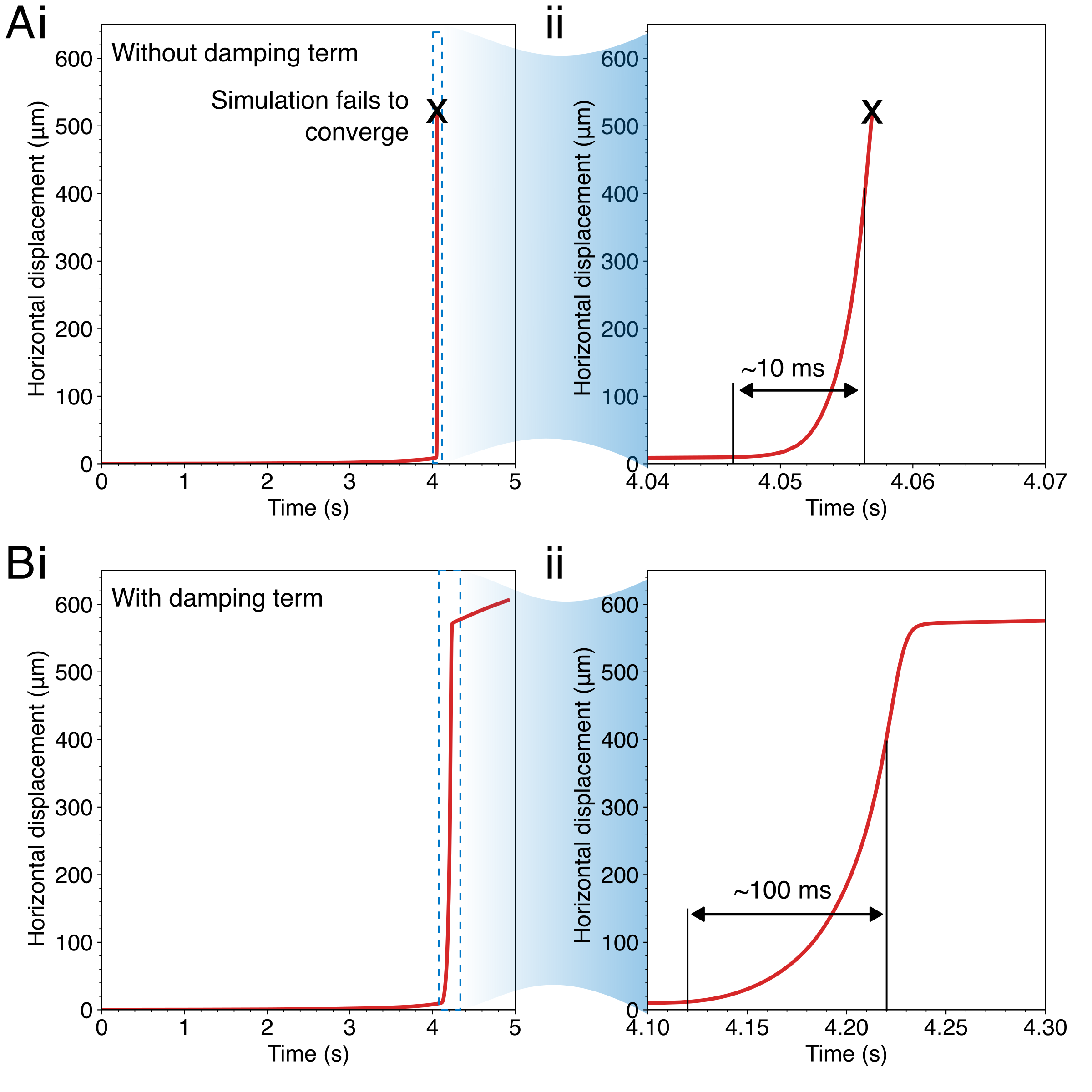}
\caption{\revision{\textbf{Numerical calculation of the 2\% dynamic range bistable bit response time.} Response times are shown (A) without and (B) with the damping term $-\eta\dot\bfu$ described in Section~\ref{app:dyn}.} }
\label{fig:SI-bistableResponseTime}
\end{centering}
\end{figure}

\clearpage 

\subsection{\revision{Summary of the coupled magneto-elasticity theory}} \label{coupled_theory}

We use the coupled magneto-viscoelasticity theory of Ref.\cite{stewart2025magnetostriction} to model the interaction between applied magnetic fields, nonlinear paramagnetic behavior of the solid material, and large elasto-dynamic deformations which drives the actuation behaviors observed in the experiments. While the theory of Ref.\cite{stewart2025magnetostriction} also includes finite-deformation viscoelasticity, here for simplicity we have neglected any viscoelastic effects. Under this simplification, the theory relates the following basic fields:\footnote{We use the standard notation of modern continuum mechanics  \citep[cf., e.g., Ref.][]{gurtin2010}.}
\begin{center}\footnotesize
\setlength{\tabcolsep}{6pt}
\renewcommand{\arraystretch}{1.1}
\begin{tabular}{ll|ll}
\multicolumn{2}{c}{\textbf{Mechanical}} &
\multicolumn{2}{c}{\textbf{Magnetic}}\\[4pt]\hline
$\bfx=\bfchi(\bfX,t)$                         & motion                                    &
$\phi(\bfX, t)$                                           & magnetostatic potential \\
    $\bfF=\nabla\bfchi,\; J=\det\bfF>0$                         &          deformation gradient                              &
$\snh$                   & magnetic field\\
       $\bfB=\F\F^\trans$     &                  left Cauchy–Green tensor     &
      $\snb$                              & magnetic flux density \\
                  $\T$             &      Cauchy stress      &
     $\mu_0 = 4\pi\times 10^{-7} \ \ \text{Tm/A}     $       & vacuum permeability \\
       $\T_{\mat}=J\T\Fit$      &          Piola stress            &
        $\snm=\mu_0^{-1}\snb-\snh$                             &  magnetization  \\
\end{tabular}
\end{center}

\noindent Since the deformation is not known a-priori, it is desirable to write the equations in the reference configuration.   The referential form of the specialized governing partial differential equations (neglecting viscoelasticity) may be summarized as follows. 

\begin{enumerate}

\item[1.]  {\bf Amp\`ere's law}:

   \begin{equation}
\Curl\snh_\mat =\zed\,, \qquad 
\text{satisfied by} \qquad 
\snh_\mat \Def - \nabla \phi \,,
        \label{pde3}
      \end{equation}

with $\snh_\mat=\F^\trans\snh $ the referential magnetic field and $\phi$ the magnetostatic potential.
\item[2.]  {\bf Gauss's law for magnetism}:

   \begin{equation}
\Div\snb_\mat =0\,, 
        \label{pde2_sum}
      \end{equation}
      
      with $\snb_\mat= J \F^\inv \snb$ the referential magnetic flux density, and 
      
\begin{equation}
 \snb  =   \underbrace{ \mu_0 \, \snh }_{\text{field response}} + \underbrace{\mu_0 \dfrac{m_s}{\sqrt{\snh\cdot\snh}}  \tanh \left(\dfrac{\chi}{m_s}\sqrt{\snh\cdot\snh}\right)  
\snh\,}_{\text{nonlinear paramagnetism}}\,, \\ 
  \label{bfieldsum2}
\end{equation}

where

\begin{itemize}
\item $\chi$  is the unitless magnetic susceptibility, and
\item $m_s$  is the saturation value of the magnetization in A/m.
\end{itemize}

\item[3.]   {\bf Equation of motion}:

   \begin{equation}
        \text{Div}\,\T_\mat = \rho_\mat \ddot\bfu\,,
        \label{pde1a_sum}
      \end{equation}

with $\T_{\mat}  =  J\,\T \F^\invtrans\,$ the Piola stress, $\rho_\mat$ the referential mass density in kg/m$^3$, and $\ddot{\bfu}$ the acceleration.  The Cauchy stress $\T$ is given by

\begin{equation}
\begin{split}
 \T =   & \underbrace{J^\inv\, \Gee \, \Bbar_0 + \Kay \, (J-1)\,\id}_{\text{mechanical stress}}  + \underbrace{\mu_0\, \left[\snh\otimes\snh  -\dfrac{1}{2} \,(\snh\cdot\snh)\id \right]}_{\text{Maxwell stress}}\, \\[4pt]
&  \underbrace{-\mu_0\,\left[ \dfrac{m_s^2}{\chi}\ln\left(\cosh\left(\chi \dfrac{\sqrt{\snh\cdot\snh}}{m_s}\right)\right)\right]\,\id  +   \mu_0\left[ \dfrac{m_s}{\sqrt{\snh\cdot\snh}}\tanh\left(\chi \dfrac{\sqrt{\snh\cdot\snh}}{m_s}\right) ( \snh\otimes\snh)\right]}_{\text{magnetization stress}}\,, \\[4pt]
 \end{split}
  \label{piolaSum2}
\end{equation}
with $\Gee$ and $\Kay$ the shear and bulk modulus, respectively, and $\bar{\bfB}_0$  the deviatoric part of the distortional left Cauchy-Green tensor $\bar\bfB=J^{-2/3}\F\FT$. 

\end{enumerate}

\subsubsection{Modeling approach for the combined fluid-solid domain}

\begin{figure}[h]
\centering
\includegraphics[width=0.35\textwidth]{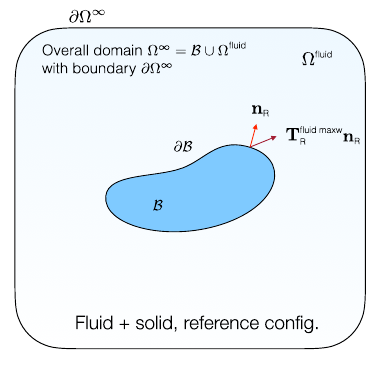}
\caption{\textbf{Referential sub-domains in the boundary-value problem in magneto-elasticity for the combined fluid-solid domain.}}
\label{domains}
\end{figure}

The magnetic fields in the fluid medium (e.g. air or water) surrounding a magnetizable solid material are crucial for determining the actuation response of the solid. We therefore define a boundary-value problem over the combined fluid-solid domain as shown in Fig.~\ref{domains}. We treat the fluid surroundings as non-magnetizable, and take the shear modulus of the fluid to be negligibly small ($G^\text{\tiny fluid}=1$\,Pa) such that the stiffness of the fluid does not impede the motion of the solid body. Finally, after Ref.\cite{stewart2025magnetostriction}, the effects of the Maxwell stress in the fluid domain on the solid are reflected by a magnetic Maxwell traction of the form
\begin{equation}
\Tmat^{\fluid \, \maxw}\bfn_\mat \Def   \underbrace{\left[J \mu_0\, \left(\snh\otimes\snh  -\dfrac{1}{2} \,(\snh\cdot\snh)\id\right)\Fit \right]}_{ \Tmat^{\text{\tiny fluid} \, \maxw}}\bfn_\mat \, \qquad \text{on} \  \partial\calB\,.
\label{tmaxwdef}
\end{equation}
Ultimately, it is the mismatch between the Maxwell traction from the fluid domain  \eqref{tmaxwdef} and the tractions from the solid domain which leads to motion of the boundary $\partial\B$. For more discussion, see Ref.\cite{stewart2025magnetostriction} and Ref.\cite{asc2025}.

\clearpage

\subsection{\revision{Summary of the modeling approach in FEniCSx}}
\label{fe_details}
The specialized governing equations for magneto-elasticity \eqref{pde3}--\eqref{piolaSum2} were solved using a fully coupled finite-element implementation in the open-source finite element software FEniCSx \citep{baratta2023, alnaes2015}, starting from existing code repositories\cite{stewart2025magnetostriction, stewart2025c, asc2025}. Three types of simulations were conducted using this numerical implementation: 
\begin{enumerate}
\item Actuation of a sphere on a thin post,
\item Actuation of a gripper device, and
\item Snap-through of a bistable bit device.
\end{enumerate}
\noindent We use meshes comprised of 2-D triangular elements which were generated using Gmsh \citep{geuzaine2009}, and we created visualizations of the results using ParaView  \citep{ahrens2005}.

 Here we summarize the major aspects of the modeling approach in FEniCSx which are not already described in Ref.\cite{stewart2025magnetostriction} and  Ref.\cite{asc2025}.  The source codes and mesh files for the FEniCSx simulations shown in this work are available from the following GitHub repository: 
\begin{itemize}
\item \url{https://github.com/ericstewart36/micro_softmagnetics}
\end{itemize}

\subsubsection{Mesh and boundary conditions}

      \begin{figure}[h]
  \centering
\includegraphics[width=\textwidth]{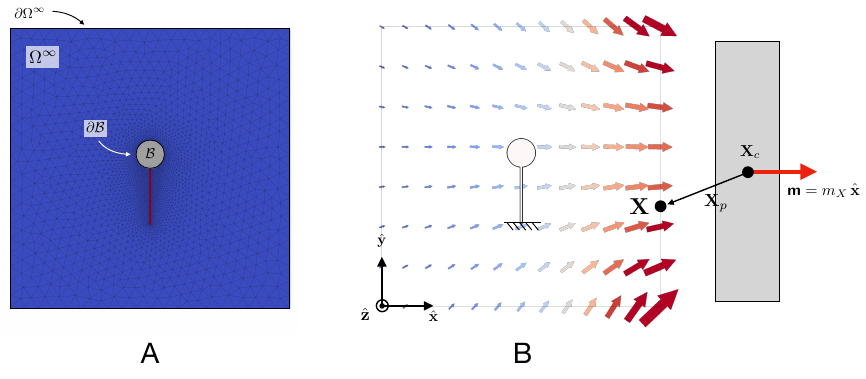}
    \caption{\textbf{Mesh and boundary conditions for actuation of a sphere on a thin post.} (A) An example mesh for the actuation of a sphere on a thin post. (B) In the simulations, the boundary $\partial\Omega^\infty$ and the bottom of the post are held fixed, while the magnetostatic potential $\phi(\X)$ set up by a bar magnet at position $\X_c$ and with magnetization $\snm=m_X\hat\x$  
   is applied to all points $\bfX$ on the boundary $\partial\Omega^\infty$. This results in a non-uniform $\snb$-field set up by the bar magnet inside the domain.  }
  \label{meshbcs}
  \end{figure}

To account for the magnetic fields in the fluid medium surrounding the magnetizable material, we construct a mesh which includes both the solid material and the surrounding fluid domain for each simulation type, Fig.~\ref{meshbcs}\,A. In the simulations of a sphere on a thin post, the boundary $\partial\Omega^\infty$ and the bottom of the post are held mechanically fixed, Fig.~\ref{meshbcs}\,B.

Next, on the boundary $\partial\Omega^\infty$ we impose the magnetostatic potential distribution $\phi(\bfX)$ generated by a rectangular prism bar magnet which:
\begin{itemize}
\item is centered at $\bfX_c = X_c \hat{\bfx} + Y_c \hat{\bfy} + Z_c\hat{\bfz}$,
\item has dimensions $T \times W \times L $ in the $\hat{\bfx}$, $\hat{\bfy}$, and $\hat{\bfz}$ directions, respectively,  and 
\item has uniform, constant magnetization $\snm=m_X\,\hat{\bfx}$. 
\end{itemize}
Denoting the relative position of point $\X = X\hat{\bfx} + Y \hat{\bfy} + Z\hat{\bfz}$  with respect to the bar magnet center as 
\[ \bfX_p \Def \bfX-\bfX_c \qquad \text{such that} \qquad X_p = (X - X_c),\ Y_p = (Y-Y_c), \ Z_p = (Z-Z_c)\,, \]
the magnetostatic potential distribution due to such a bar magnet is given by (\cite[cf., e.g., Ref.][]{james2026}),
\begin{equation}\label{barPot}
\begin{split}
\phi(\bfX) = \hat\phi(X_p,Y_p,Z_p) = \dfrac{m_x}{4\,\pi} \, \Bigg[  &G\bigg(X_p-T/2 \bigg)  - G\bigg(X_p+T/2\bigg)\Bigg] \,,
\end{split}
\end{equation}
where
\begin{equation}
\begin{split}
G(\xi) \Def  \Big[ &F\bigg(\xi, Y_p- W/2,  Z_p -  L/2 \bigg) - F\bigg(\xi, Y_p- W/2,  Z_p +  L/2 \bigg) \Big] \\
- \Big[ &F\bigg(\xi, Y_p+ W/2,  Z_p -  L/2 \bigg) - F\bigg(\xi, Y_p+ W/2,  Z_p +  L/2 \bigg) \Big] \,,
\end{split}
\end{equation}
and
\begin{equation}
\begin{split}
&F(i,j, k) \Def -i \tan^\inv \left(\dfrac{jk}{ir}\right) + j\ln(k+r) + k\ln(j+r)\, \\
&\text{with} \qquad r \Def \sqrt{i^2 + j^2 + k^2}\,. \\
\end{split}
\end{equation}

\noindent 

The bar magnet center position is then altered as a function of time, $\bfX_c = \hat\bfX_c(t)$, in order to prescribe the motion of the bar magnet relative to the magnetizable material.
Finally, to account for the effects of Maxwell stresses in the fluid surroundings, Maxwell tractions of the form \eqref{tmaxwdef} are applied to the boundary $\partial\B$ at the interface between the magnetizable solid domain and the fluid domain.\footnote{For simplicity, in the sphere on a thin post and gripper actuation studies  the magnetization of the support posts are neglected. Numerical studies showed that this treatment does not significantly affect the results. Accordingly, in these simulations the boundary $\partial\calB$ to which Maxwell tractions are applied is taken to be the boundary of the magnetizable circular domains at the end of each post.}

Table~\ref{barParams} lists the dimensions and magnetization strength of the bar magnets considered in this work, which are key inputs to the form of the magnetostatic potential in \eqref{barPot}. This table also lists the magnitude of the $\snb$-field gradient in the vicinity of each bar magnet.

  \begin{table}[h]
\centering
\caption{Characteristics of the different bar magnets.}
\begin{tabular}{cccccc}
\hline
Bar magnet & Field gradient & $T$ & $W$ & $L$ & $m_X$ \\
                    & (mT/mm) & (mm) &  (mm) & (mm)  & (MA/m) \\
\hline
BX041-N52 & 30 & 1.574 &  6.35 & 25.4 & 1.178  \\
BX042-N52 & 80 & 3.17 &  6.35 & 25.4 & 1.178  \\
BX044-N52 & 120 & 6.35 &  6.35 & 25.4 & 1.178  \\
\hline
\end{tabular}
\label{barParams}
\end{table}

\noindent  The 30 mT/mm bar magnet (BX041-N52) was used for the simulations of the actuation of a sphere on a thin post, while the 120 mT/mm bar magnet (BX044-N52) was used for the simulations of the gripper and the bistable bit. We omit a full description of the boundary conditions for the magnetic gripper and bistable bit simulations, as they are similar to those described for the actuated sphere on a thin post.

  \subsubsection{Material parameters}
  
 Table~\ref{params} lists the material parameters for the four different dynamic range values (and thus four different cross-linking densities) which are considered in the simulations. Throughout, we assume that the polymer gels have a mass density $\rho_\mat=1000$\,kg/m$^3$ and are nearly incompressible such that $K/G=10^3$\, and $G$ is well-approximated by $E/3\,$.
  
  \begin{table}[h]
\centering
\caption{Material properties used in the simulations for different dynamic range values.}
\begin{tabular}{ccccc}
\hline
\text{Dynamic Range} & $G$ & $K$ & $\chi$ & $m_s$ \\
\text{(\%)} & \text{(kPa)} & \text{(kPa)} & -  & \text{(MA/m)} \\
\hline
2  & 6.652$\times10^3$ & 6.652$\times10^6$ & 0.23 & 0.00466688 \\
18 & 12.010$\times10^3$ & 12.010$\times10^6$ & 0.23 & 0.00180904 \\
51 & 17.945$\times10^3$ & 17.945$\times10^6$ & 0.23 & 0.0009868  \\
100 & 17.734$\times10^3$ & 17.734$\times10^6$ & 0.23 & 0.00054656 \\
\hline
\end{tabular}
\label{params}
\end{table}

\subsubsection*{Plane strain formulation}

For simplicity, all three simulation types use a 2-D plane strain formulation. The bistable bit experiment itself is plane strain. However, for the actuation of a sphere on a thin post and the gripper simulations there are two deviations from plane strain conditions: (i) the thin post has a cylindrical cross-section, and not a square cross-section, and (ii) the sphere diameters are wider than the posts' out-of-plane thickness. We address these items using the adjustment factors described below. We emphasize that
\begin{itemize}
\item these adjustment factors are approximative in nature, and therefore the simulation results for the actuation of a sphere on a thin post and the gripper devices are not precise quantitative predictions of the corresponding experiments,
\end{itemize}
however the simulations still capture the essential coupled-physics phenomena underlying the experiments and predict the variation of the actuation behavior with changes in geometry, boundary conditions, and material parameters.\footnote{Full 3-D simulations of the coupled magneto-elasto-dynamics of the spherical actuators and the surrounding fluid domain are computationally expensive and exhibit significant numerical convergence challenges; we leave such studies to future work. }

\noindent \textbf{Cylindrical post cross-section adjustment:}
While the actual 3D post's cross-section is a circle with diameter $t=33.3\,\mu$m, the 2D plane strain (PE) post cross-section is a square with side length $t=33.3\,\mu$m. The corresponding area moments of inertia are
\begin{equation}
I_\text{PE} = \dfrac{1}{12}t^4 \, \qquad \text{and} \qquad I_\text{3D} = \dfrac{1}{64}\pi t^4 \,.
\end{equation}
We require that the plane-strain beam have the same flexural rigidity as the 3D cylindrical beam, 
\begin{equation}
(EI)_\text{PE} = (EI)_\text{3D} \qquad \text{which requires} \qquad E_\text{PE} = \bigg( \dfrac{3\pi}{16} \bigg) \, E \approx 0.59 \, E  \,.
\end{equation}
This correction factor of $(3\pi)/16$ is applied to the stiffness of the thin post in both the sphere on a thin post and gripper actuation simulations. 

\noindent \textbf{Sphere volume adjustment:}
The plane strain formulation treats all domains as having a uniform out-of-plane thickness $t=33.3\,\mu$m, but the sphere has a radius $r>33.3\,\mu$m. The ratio of the actual sphere volume to the plane strain ``ball'' volume is
\begin{equation}
V_\text{fac} = \dfrac{V_\text{sphere}}{V_\text{PE}} = \dfrac{4/3 \, \pi r^3}{ \pi r^2 t} = \bigg( \dfrac{4}{3}\bigg) \bigg( \dfrac{r}{t} \bigg)\,.
\end{equation}
We adjust the mass density of the plane strain ``ball'' according to $\hat{\rho}_\mat \Def V_\text{fac} \, \rho_\mat$ so that the total mass of the plane strain ``ball'' is the same as the total mass of the 3-D sphere.

Then, we can approximate the total magnetization of a 3D sphere as $M_\text{tot}^\text{sphere} = \vert \snm \vert \, V_\text{sphere}$ and
 the total magnetization of the plane strain ``ball'' as $M_\text{tot}^\text{PE} =  \vert\snm \vert\, V_\text{PE}.$
We therefore adjust the magnetization response  of the plane strain ``ball''  as $\hat{\snm}(\snh) = V_\text{fac} \, \snm(\snh)\,,$
such that the total magnetization in the simulation matches the total magnetization of the 3D sphere:
\begin{equation}
 M_\text{tot}  = V_\text{fac} \, \vert \snm \vert V_\text{PE} = \vert \snm \vert \, V_\text{sphere} = M_\text{tot}^\text{sphere}\,. 
\end{equation}
To this end we take $\hat{m}_s = V_\text{fac} \, m_s$, and --- since $\snm=\chi\snh$ in the limit of small $\snh$ --- we also take $\hat{\chi} = V_\text{fac} \, \chi$.

\subsubsection{Treatment of dynamic effects}
\label{app:dyn}

At each time step, we compute the velocity and acceleration  using the trapezoidal Newmark time-integration scheme     \citep{newmark1959}:

\begin{equation}
 \begin{split}
\ddot\bfu &= \dfrac{ \bfu - \bfu_\old- \dot\bfu_\old\,\Delta t}{(1/4) \,  \Delta t^2} - \ddot\bfu_\old  \,,\\[4pt]
 \dot\bfu &=  \dot\bfu_\old +\frac{1}{2} \, \Delta t \,\left( \ddot\bfu_\old + \ddot\bfu\right)\,.
 \end{split}
\label{pdeweaksum2b}
\end{equation}
 
 The acceleration $\ddot{\bfu}$ is used to compute the inertial term $\rho_\mat\ddot{\bfu}$ in the mechanical governing equation \eqref{pde1a_sum}. 
Iteration-based adaptive time-stepping proved necessary for convergence during the onset of instabilities and during high-acceleration events, such as the snap-through instability in the bistable bit simulation.

In the simulations we also include a viscous body force such that the mechanical governing equation \eqref{pde1a_sum} becomes 
   \begin{equation}
        \text{Div}\,\T_\mat + \bff_\text{\mat,visc} = \rho_\mat \ddot\bfu\, \qquad \text{with} \qquad \bff_\text{\mat,visc} = -\eta\dot\bfu \,,
        \label{pde1a_sum_b}
      \end{equation}

which accounts for the hydrodynamic ``drag forces'' due to the motion of the solid through the surrounding fluid medium. This viscous term is needed so that the dynamic oscillations of the actuators eventually settle during motion, rather than continuing to oscillate for all time. It also provides numerical damping which is useful for convergence. A value of $\eta = 10^{-3}$ kg/(mm$^3$s) was used in the sphere on a thin post and gripper simulations, and a value of  $\eta = 10^{-2}$ kg/(mm$^3$s) was used for the bistable bit simulation. The effect of this numerical damping term on the response time and convergence behavior of the bistable bit simulation can be seen in Figure~\ref{fig:SI-bistableResponseTime}.

\clearpage

\bibliography{references}

\end{document}